\documentclass[twocolumn]{aastex63}
\usepackage{natbib}
\usepackage{amsmath} % or simply amstext

\newcommand{\msun}{$M_{\odot}$}
\newcommand{\sersic}{S\'ersic}

\begin{document}

\title{The Nature of Low Surface Brightness Galaxies in the Hyper Suprime-Cam Survey}
\author[0000-0002-5612-3427]{Jenny E. Greene}
\affiliation{Department of Astrophysical Sciences, Princeton University,Princeton, NJ 08544, USA}
\author{Johnny P. Greco}
\affiliation{Center for Cosmology and AstroParticle Physics (CCAPP), The Ohio State University, Columbus, OH 43210, USA}
\author[0000-0003-4700-663X]{Andy D. Goulding}
\affiliation{Department of Astrophysical Sciences, Princeton University,Princeton, NJ 08544, USA}
\author[0000-0003-1385-7591]{Song Huang}
\affiliation{Department of Astrophysical Sciences, Princeton University,Princeton, NJ 08544, USA}
\affiliation{Department of Astronomy, Tsinghua University, Beijing 100084, China}
\author[0000-0002-0332-177X]{Erin Kado-Fong}
\affiliation{Department of Astrophysical Sciences, Princeton University,Princeton, NJ 08544, USA}
\author[0000-0002-1841-2252]{Shany Danieli}
\altaffiliation{NASA Hubble Fellow}
\affiliation{Department of Astrophysical Sciences, Princeton University,Princeton, NJ 08544, USA}
\author[0000-0001-9592-4190]{Jiaxuan Li}
\affiliation{Department of Astrophysical Sciences, Princeton University,Princeton, NJ 08544, USA}
\author[0000-0002-1418-3309]{Ji Hoon Kim}
\affiliation{Subaru Telescope, National Astronomical Observatory of Japan, National Institutes of Natural Sciences (NINS), 650 North Aohoku
Place, Hilo, HI 96720, USA}
\affiliation{SNU Astronomy Research Center, Seoul National University, 1 Gwanak-ro, Gwanak-gu, Seoul 08826, Republic of Korea}
\author[0000-0002-3852-6329]{Yutaka Komiyama}
\affiliation{Dept. of Advanced Sciences, Faculty of Science and Engineering, Hosei University,
3-7-2 Kajino-cho, Koganei-shi, Tokyo 184-8584, Japan}
\author[0000-0002-3677-3617]{Alexie Leauthaud}
\affiliation{Department of Astronomy and Astrophysics, University of California Santa Cruz, 1156 High St., Santa Cruz, CA 95064, USA}
\author{Lauren A. MacArthur}
\affiliation{Department of Astrophysical Sciences, Princeton University,Princeton, NJ 08544, USA}
\author[0000-0002-8149-1352]{Crist\'obal Sif\'on}
\affiliation{Instituto de F\'isica, Pontificia Universidad Cat\'olica de Valpara\'iso, Casilla 4059, Valpara\'iso, Chile}

\date{December 2021}

\begin{abstract}
    We present the statistical redshift distribution of a large sample of low surface brightness (LSB) galaxies identified in the first 200 deg$^2$ of the Hyper Suprime-Cam Strategic Survey Program. Through cross-correlation with the NASA-SDSS Atlas, we find that the majority of objects lie within $z<0.15$ or $\sim 500$~Mpc, yielding a mass range of $M_* \approx 10^7-10^9$~\msun\ and size range of $r_{\rm eff, g} \approx 1-8$~kpc. We find a peak in the distance distribution within 100 Mpc, corresponding mostly to $\sim 10^7$~\msun\ galaxies that fall on the known mass-size relation. There is also a tail in the redshift distribution out to $z \approx 0.15$, comprising more massive ($M_*=10^8-10^9$~\msun) galaxies at the larger end of our size range. We see tentative evidence that at the higher-mass end ($M_* > 10^8$~\msun) the LSB galaxies do not form a smooth extension of the mass-size relation of higher surface brightness galaxies, perhaps suggesting that the LSB galaxy population is distinct in its formation path. 
\end{abstract}

\section{Introduction}

Low surface brightness (LSB) galaxies have long tantalized astronomers \citep[LSB galaxies; e.g.,][]{Sandage:1976aa,CaldwellBothun:1987aa,Impey:1988aa,Bothun:1991aa,McGaugh:1995ab,Dalcanton:1997aa}. A significant fraction of galaxies could be LSB \citep[e.g.,][]{Dalcanton:1997ab}. If galaxies are selected by surface brightness alone, there is a wide range of mass and morphology that can result. Specifically, in this work we focus on galaxies with $M_*<10^{9.5}$~\msun\ that have no bulges, as opposed to so-called ``giant'' LSB disk galaxies at higher mass \citep[such as the prototype Malin 1][]{Bothun:1987aa}.

These low-mass LSB galaxies have been of interest because they may result from the high angular momentum tail of the halo distribution \citep[e.g.,][]{Dalcanton:1997aa,Amorisco:2016aa}, or extreme feedback events \citep[e.g.,][]{elBadry:2016aa,DiCintio:2017aa,Chan:2018}. Others have suggested that it is interactions that make galaxies larger, either galaxy merging at early times \citep[e.g.,][]{Wright:2021aa}, falling into the cluster environment \citep{van-Dokkum:2015aa,van-Dokkum:2015ab,Safarzadeh:2017aa}, tidal effects \citep{Iodice:2021aa,Benavides:2021aa}, or (very likely) multiple paths \citep[e.g.,][]{Papastergis:2017aa}. Interestingly, the intrinsic shapes of LSB galaxies appear to be thicker than ``normal'' dwarfs at matched stellar mass, and this shape difference appears similar across environments \citep{Kado-Fong:2021aa}.

Much of the recent excitement to revisit this population has been driven by improvements in available imaging, with wide-format CCDs on small \citep[e.g.,][]{Abraham:2014aa,Martinez-Delgado:2016aa,Carlsten:2020aa} and large \citep[e.g.,][]{Aihara:2018aa,Dey:2019aa} aperture telescopes generating wide, deep, and high-resolution imaging data sets. Very LSB galaxies reaching central surface brightness limits of 26-29 mag~arcsec$^{-2}$ are being uncovered in unprecedented numbers by modern surveys. Two types of searches are worth highlighting here. A recent paper that reignited excitement around LSB galaxy searches identified a very large number of such galaxies around the Coma cluster \citep{van-Dokkum:2015aa}. While prior work had identified isolated examples of such galaxies \citep{Sandage:1976aa}, the exquisite surface brightness sensitivity of the Dragonfly telescope array revealed a large population of LSB galaxies in cluster environments, which have since been studied further \citep[e.g.,][]{Koda:2015aa,Mihos:2015aa,Munoz:2015aa,Yagi:2016aa,van-der-Burg:2016aa,Martinez-Delgado:2016aa,Lee:2017} including down to groups \citep[][]{van-der-Burg:2017aa} and the field \citep[e.g.,][]{Roman:2017aa,Leisman:2017aa,Greco:2018ab,Roman:2019}. 

The benefit of focusing on rich environments is the ability to assume a distance, and thus measure physical properties. However, we are then left wondering whether the cluster environment is a requirement to make such extended systems. Blind searches across wide areas are complementary, and in this paper we focus on deriving the distance distribution for one of the first blind modern searches for LSB galaxies selected in the stellar continuum, \citet[][G18 hereafter]{Greco:2018aa}. There are other larger-area searches \citep[][]{Zaritsky:2019aa,Tanoglidis:2021aa,Zaritsky:2021aa} but they do not reach the surface brightness sensitivity of G18 (\S \ref{sec:complete}). Radial velocities for individual targets also provide robust physical properties \citep[e.g.,][]{van-Dokkum:2016aa,Greco:2018ab,Kadowaki:2021aa}, but are quite expensive and grow prohibitive at the lower surface brightness end of the samples. A statistical approach to estimate the distance distribution can complement follow-up of individual objects.

Our goal here is to characterize the distance, mass, and size distribution of the galaxies presented in G18. We follow \citet{Menard:2013aa} and use spatial cross-correlation to constrain the redshift distribution statistically. Throughout the paper, we assume a hubble constant $H_0 = 70$~km~s$^{-1}$~Mpc$^{-1}$, $\Omega_M = 0.3$, $\Omega_{\Lambda} = 0.7$.

%%%%%%%%%%%%%%%%%%%%%%%%%%%%%%%%%%
\begin{figure}
\includegraphics[width=0.50\textwidth]{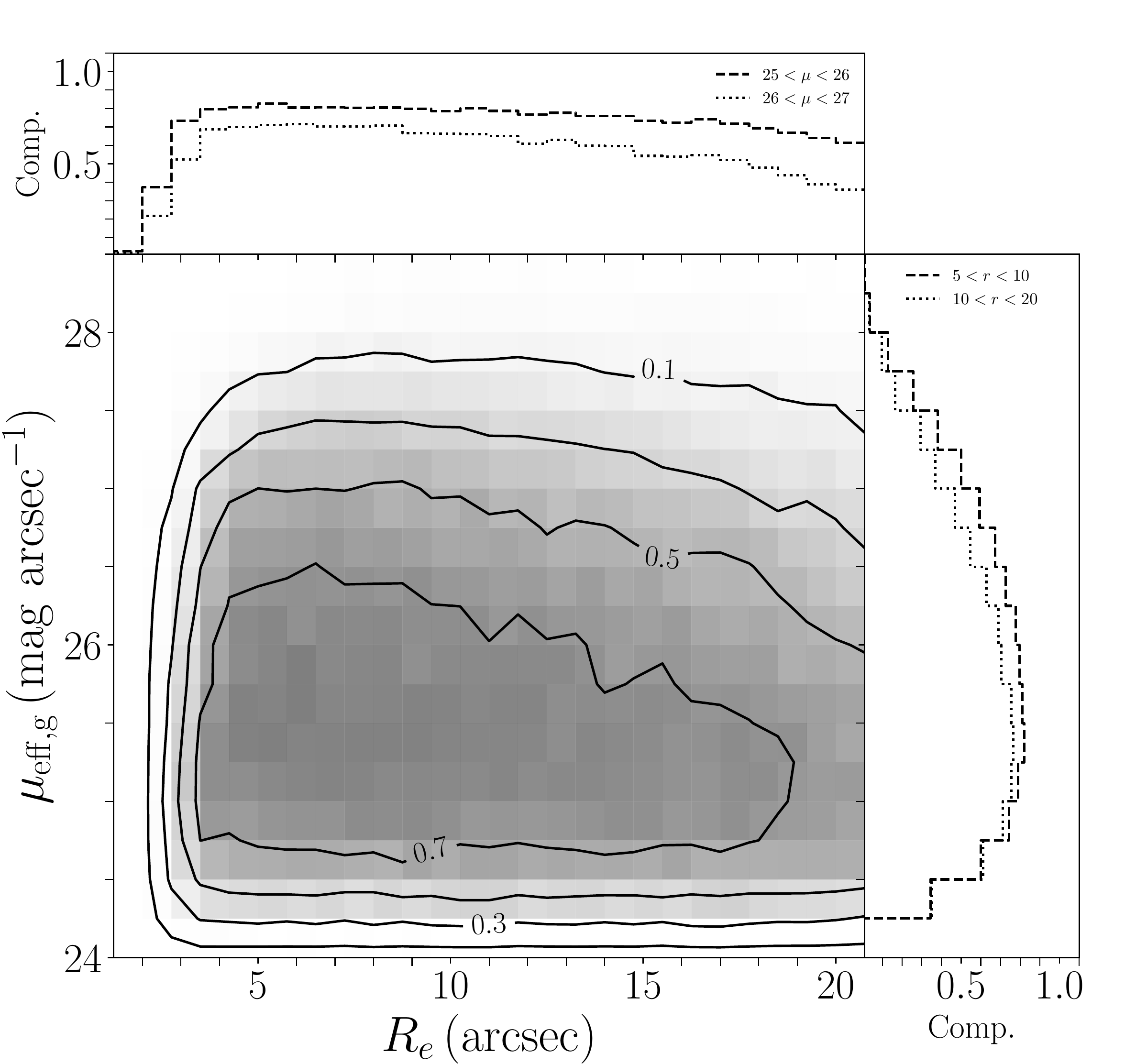}
\caption{Completeness of the G18 sample as a function of size and surface brightness. Detailed image simulations were run to define these completeness levels. The completeness is a weak function of size (top histogram) and a stronger function of surface brightness (side histogram).
\label{fig:complete}}
\end{figure}
%%%%%%%%%%%%%%%%%%%%%%%%%%%%%%%%%%

\section{Samples}
\label{sec:sample}

\subsection{A Search with Hyper Suprime-Camera}
\label{sec:search}

%HSC paragraphs here.
The Hyper Suprime-Camera \citep[HSC;][]{Miyazaki:2018aa,Komiyama:2018aa} on the Subaru Telescope provides one of the largest fields of view on an 8m-class telescope. The HSC Strategic Survey Program (HSC-SSP) is a $\sim 5$ year survey using HSC in five bands ($grizY$) \citep[see][for an overview]{Aihara:2018aa}. The Wide layer ($\sim 1000$ deg$^2$ to $i=25.2$ AB mag $5 \, \sigma$ point source detection limit) has weak lensing cosmology as its primary focus \citep[e.g.,][]{Hamana:2020aa,Miyatake:2021aa}, and is also the basis for this paper. The 200 deg$^2$ analyzed by G18 fall within the Public Data Release 2 \citep{Aihara:2018ab} area. The data are processed using \emph{hscPipe} \citep{Bosch:2018aa}. 

%%%%%%%%%%%%%%%%%%%%%%%%%%%%%%%%%%
\begin{figure}
\includegraphics[width=0.50\textwidth]{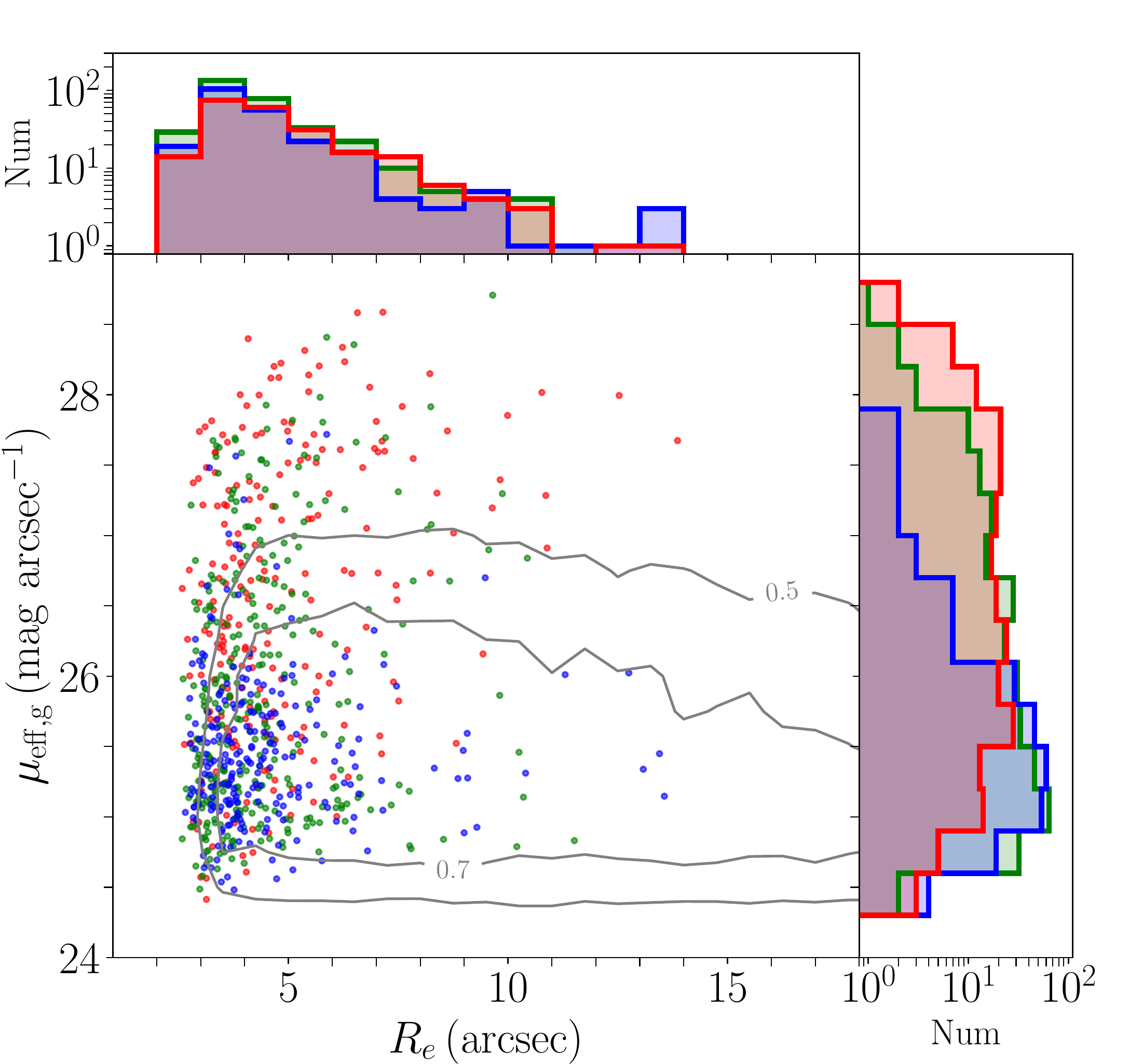}
\caption{Distribution in size and surface brightness of the G18 sample. The surface brightness histogram reveals that the blue sample ($g-i<0.5$) spans a much narrower range in $\mu_{\rm eff, g}$ than the red ($g-i>0.8$) or green ($0.5<g-i<0.8$) samples. The galaxies of intermediate color fall between the red and blue samples in their surface brightness distributions. The 50\% and 70\% completeness contours are shown.
\label{fig:sbcolor}}
\end{figure}
%%%%%%%%%%%%%%%%%%%%%%%%%%%%%%%%%%

G18 performed an LSB galaxy search with the first $\sim 200$~deg$^2$ of HSC data. The details of their search are described in \S2 of G18, but for completeness we summarize the main elements of their search here. First, each image is smoothed with a Gaussian kernel that matches the point spread function (PSF) with $i \approx 0.6$\arcsec. Bright objects are selected as groups of pixels that are $28 \sigma$ above the background level. LSB structures are flagged as groups of at least 20 pixels that are $3 \sigma$ above the background, and these are linked to bright objects if at least $15\%$ of the pixels in an LSB structure are above the threshold for the bright object, with these values chosen by trial and error. With images clean of bright sources and their LSB outskirts, G18 then identify LSB candidates in the $i-$band images using {\tt SExtractor} \citep{Bertin:1996aa}. The images are convolved with a $1\arcsec$ kernel, roughly twice the PSF. Larger kernels boost sensitivity but introduce blends of faint background sources as a major contaminant. The local background is remeasured on 22\arcsec\ scales, and groups of 100 total pixels (corresponding to radii of $\sim $roughly 1-2 times the PSF) with at least $0.7 \sigma$ above the background are chosen as potential sources. Aperture photometry for selection only is performed with an aperture that is 2.5 times the Kron radius, with forced photometry performed in the $g,\,r$ bands for color information. The search is performed in independent $12\arcmin \times 12$\arcmin\ patches.

Initial candidates are then selected using size and color cuts. Specifically, the {\tt FLUX\_RADIUS} parameter from {\tt SExtractor} is restricted to $>2.5$\arcsec, since G18 are looking for extended LSB galaxies. This cut alone reduces the candidate list by two orders of magnitude. At the large end, G18 are limited by the patch overlap regions (17\arcsec) and larger sizes are unreliable, so G18 impose an upper size cut of 20\arcsec. At this stage, G18 also use the forced photometry to remove very red galaxies. Tens of thousands of candidates remain after this step. Finally, two-dimensional \sersic\ modeling is done using {\it imfit} \citep{Erwin:2015aa}. The final colors are based on a $g-$band model fixed to the structural parameters from the $i-$band fit.  After this step, the $\mu_{\rm g,eff} > 24.3$~mag~arcsec$^{-2}$ and $r_{\rm eff} > 2.5$\arcsec\ cuts are applied, leaving 1521 candidates. Two coauthors visually inspect the candidates to remove tidal features, cirrus, and clusters of point sources that got through the above cuts. Undoubtably, some incompleteness is introduced with this step that is not quantified below. The final candidate list contains 781 galaxies.

\subsection{Sample Properties}

The G18 completeness (\S \ref{sec:complete}) is shown in Figure \ref{fig:complete}, which leads to a distribution in size and surface brightness as shown in Figure \ref{fig:sbcolor}. The sample has a wide range in optical color, with a median of $\langle g-i \rangle = 0.64$. The red galaxies (taking the median of all galaxies redder than $g-i = 0.64$ yields $\langle g-i \rangle = 0.8$) show smooth and round morphologies and span the full range of surface brightness detected in the sample. The blue galaxies (median of the blue galaxies is $\langle g-i \rangle = 0.5$) are more irregular in shape but span a narrower range of surface brightness ($\mu_{\rm eff, g} \lesssim 25.5$ AB mag arcsec$^{-2}$). The blue LSB galaxies are actively forming stars, as evidenced by their near-ubiquitous detection with \emph{GALEX}, as well as detected H$\alpha$ in a handful of galaxies for which we have spectroscopic follow-up \citep{Greco:2018ab}. \citet{Kado-Fong:2021aa} investigate the intrinsic shapes of the LSB galaxies in this sample, along with an \ion{H}{1}-selected sample \citep{Leisman:2017aa}. They match the mass ranges of the \ion{H}{1}-selected and ``normal'' galaxies, finding that the LSB galaxies have intrinsic axis ratios $C/A$ that are $\sim 1.5-2$ times larger than the normal dwarfs at matched stellar mass. With a mass distribution from this analysis, we can check whether this shape holds robustly for optically selected samples like this one. 

Thus far, we have distances for only a handful of galaxies in the G18 sample, from radial velocities \citep{Greco:2018ab} or indirectly from globular cluster luminosity functions \citep{Somalwar:2020aa}. Deeper investigations into the nature of these galaxies requires knowing their distances (at least statistically), which is the goal of the present work. Before we proceed with the cross correlation analysis, we first need to know the sample completeness, presented in the next subsection.

\subsection{Completeness}
\label{sec:complete}

We have performed image simulations to understand our completeness as a function of galaxy size and surface brightness. We inject $\sim 350,000$ artificial galaxies with a range of true average surface brightness of $23 \leqslant \mu_{\rm eff, g} \leqslant 28.5$~mag~arcsec$^{-2}$. The simulated galaxies are modeled as single \sersic\ functions, with $2\arcsec \leqslant r_e \leqslant 21\arcsec$, $0.8 < n < 1.2$, and $0 < \epsilon < 0.6$ to match the G18 distribution. They are injected into the coadded final frames and recovered using the software presented in G18. 

The completeness as a function of size and surface brightness based on these tests is shown in Figure \ref{fig:complete}. The completeness falls off towards higher surface brightness because the sample was selected to be fainter than $\mu_{\rm eff, g}>24.3$~mag~arcsec$^{-2}$. While nominally we selected galaxies larger than $r_{\rm eff} > 2.5$\arcsec, we are most complete for somewhat larger sizes, $r_{\rm eff} > 4$\arcsec, likely due to blending of smaller sources with the background that makes it hard for the search algorithm to identify them as unique sources. At larger size, however, the completeness is not very sensitive to size until the very largest sizes ($r_{\rm eff} > 15$~\arcsec) where a mild decrease is seen. We are $\sim 80\%$ complete to $\mu_{\rm eff, g} < 26.5$~mag for $r_{\rm eff} < 10$\arcsec. In this size range, we are still $\sim 70\%$ complete at $\mu_{\rm eff, g} = 28.5$~mag~arcsec$^{-2}$. For larger sizes, we are slightly less complete with surface brightness, reaching a maximum completeness of $\sim 70\%$ for $\mu_{\rm eff, g} < 25.5$~mag, and dropping to $\sim 50\%$ at $\mu_{\rm eff, g} = 28.5$~mag~arcsec$^{-2}$. This size dependence is due in part to the search being performed in individual HSC patches as discussed in \S \ref{sec:search}. 

We also perform a smaller suite of tests in which the galaxies are injected prior to sky subtraction. In these simulations, we track the recovery of key parameters such as \sersic\ index, ellipticity, and color. We also test a wider range of \sersic{} indices $0.3<n<2.5$. These tests are detailed extensively in \citet{Greco:2018phd}. Here, we summarize that while we see no dependence of completeness on \sersic\ index, there is a bias in the fits that prefers lower values of $n<1.5$, even when the injected value is higher \citep[see Figure 5.5 in][]{Greco:2018phd}. We also find a rapidly declining completeness as a function of ellipticity above $\epsilon \approx 0.6$ due to surface brightness projection effects, suggesting that some edge-on disk galaxies may be missing from the sample \citep[Figure 5.8; see also discussion in][]{Kado-Fong:2021aa}. We find no color dependence in the completeness over the range of $g-i$ color observed in the sample, nor any bias in the color recovery. Finally, \citet{Greco:2018phd} also test the impact of adding additional structure (e.g., star-forming clumps) to the blue model galaxies and also find negligible impact to the recovery fractions.

These completeness levels highlight the power of the HSC survey (and soon the Vera Rubin Observatory, \citealt{Ivezic:2019}) to perform sensitive searches in the low surface brightness regime over wide areas and thus a wide range of environments. For context, earlier searches that quantified the number density of LSBs \citep[e.g.,][]{Dalcanton:1997aa} had a maximum completeness of $\sim 40\%$ at $23 < \mu_0 < 25.5$ mag arcsec$^{-2}$. Like our search, their completeness was very poor for the smallest galaxy sizes $<2$\arcsec. \citet{van-der-Burg:2017aa} search for LSB galaxies in the ESO Kilo-Degree Survey, and achieve $\sim 80\%$ completeness at $\mu_{\rm eff, r} \approx 25.5$~mag~arcsec$^{-2}$, or roughly $\mu_{\rm eff, g} \approx 26.1$~mag. \citet{Carlsten:2020aa} used a very similar algorithm to ours, and with the Canada France Hawai'i Telescope Legacy Survey data were able to reach $\mu_0 \sim 26.5$~mag arcsec$^{-2}$ at $\sim 50\%$ completeness, which is roughly similar to $\mu_{\rm eff} \sim 25.8$~mag arcsec$^{-2}$ for a \sersic{} index of $n=1$ \citep[e.g.,][]{Greco:2018aa}. Finally, \citet{Tanoglidis:2021aa} performed a wide-area search using the Dark Energy Survey (DES) images \citep{DES:2018aa}. \citet{Kado-Fong:2021aa} compare the surface brightness distributions of the G18 and Tanoglidis et al.\ samples, to show that they are complete at a similar $\sim 70-80\%$ level to roughly $\mu_{\rm eff, g} = 25.75$~mag~arcsec$^{-2}$, and this value is comparable to estimates from Tanoglidis et al.\ Dragonfly routinely reaches $\mu_r \sim 29.5$~mag~arcsec$^{-2}$ at $1 \, \sigma$ on 5\arcsec\ scales, so such limits are in reach for large volume searches albeit at lower spatial resolution \citep[][]{Danieli:2020aa}. Thus, the combination of depth, area, and resolution makes the HSC SSP survey data an excellent tool for LSB galaxy searches \citep[e.g.,][]{Huang:2018aa,Kado-Fong:2020aa}.

\subsection{Volume corrections}

Because of the size and surface brightness cuts we impose, we will not be equally sensitive to all galaxies in the sample throughout the full volume. In the standard way, we correct for this incompleteness by calculating the maximum volume out to which each source is detectable, and then weight each source by both the detection completeness at that size and surface brightness, but also its volume completeness. 

In practice, it is not the magnitude that determines when a galaxy falls out of the sample in distance, but rather the size and surface brightness.  To calculate the maximum volume, we include two terms. The first is the maximum distance at which the galaxy is large enough to remain in our sample ($r_e > 2.5$\arcsec). The second is the maximum distance at which the galaxy has $\mu_{\rm eff} < 27$ mag~arcsec$^{-2}$ (our approximate $50\%$ limit). Whichever volume is smaller sets the $V_{\rm max}$ weighting for that galaxy. 

\subsection{Redshift reference samples}

In addition to our primary sample of LSB galaxies with an unknown distance distribution, we will rely on a reference sample with known redshifts to cross-correlate with our sample. The ideal reference sample includes very nearby galaxies ($\sim 20$~Mpc) and extends to at least redshift $z=0.15$ (700 Mpc). In addition, the spatial sampling of the reference sample should be well-understood, such that we can identify regions of the sky where no spectra could have been taken due to issues like bright stars, incomplete imaging coverage, or other non-galaxy--related issues. 

We considered using the Extragalactic Distance Database \citep{Tully:2009aa}, but that sample starts to become incomplete above $\sim 200$~Mpc, and thus does not cover the full distance distribution of our sample (as we justify in \S \ref{sec:crosscorr}). Similar concerns hold for the Two Mass Redshift Survey \citep{Kochanek:2001aa}. The Galaxy and Mass Assembly (GAMA) survey \citep{Driver:2011aa} only covers a small fraction of our areal coverage. Therefore, we have relied on two complementary samples that are both built from the Sloan Digital Sky Survey \citep[SDSS;][]{York:2000aa}. As our primary reference sample, we use the NASA-Sloan Atlas\footnote{In this work, we use Version 1\_0\_1: \url{https://www.sdss.org/dr13/manga/manga-target-selection/nsa/}} \citep[NSA;][]{Blanton:2005aa,Blanton:2011aa}. This is a catalog that painstakingly combines the redshift survey of the SDSS Legacy galaxy sample with a magnitude limit of $r<17.77$~AB mag \citep{Strauss:2002aa} and literature redshifts to fill in the range below $z \approx 0.02$. Those redshifts and distances may come from gas, direct distance measurements, or anything else available in the literature, and therefore the spatial distribution of this sample is not well known, so we cannot construct a random catalog for the NSA sample. 

To complement the NSA tracer sample, therefore, we also use the SDSS Large Scale Structure catalog from Data Release 12 \citep[][]{Ross:2011aa,Ho:2012aa,Reid:2016aa}. This sample is a combination of the Legacy sample referenced above and the BOSS spectroscopic survey \citep{Dawson:2013aa}. The spatial sampling of these redshifts is very well-understood and a catalog of random positions is provided \citep{Alam:2017aa}. There are two reasons that we do not use the SDSS DR12 sample as our reference. First, the catalog only contains galaxies with $z>0.02$, and many of our targets are closer than that \citep{Greco:2018ab}. Second, because of the coverage of the catalog, we lose roughly a quarter of our areal coverage where there are no objects in the SDSS DR12 catalog. Thus, this sample serves mostly as a check that we recover the same redshift distribution when we are controlling the spatial distribution of the tracer sample.

\section{The LSB redshift distribution}
\label{sec:crosscorr}

%%%%%%%%%%%%%%%%%%%%%%%%%%%%%%%%%%
\begin{figure*}
\centering
\begin{minipage}{.98\textwidth}
    \centering
    \includegraphics[width=0.45\linewidth]{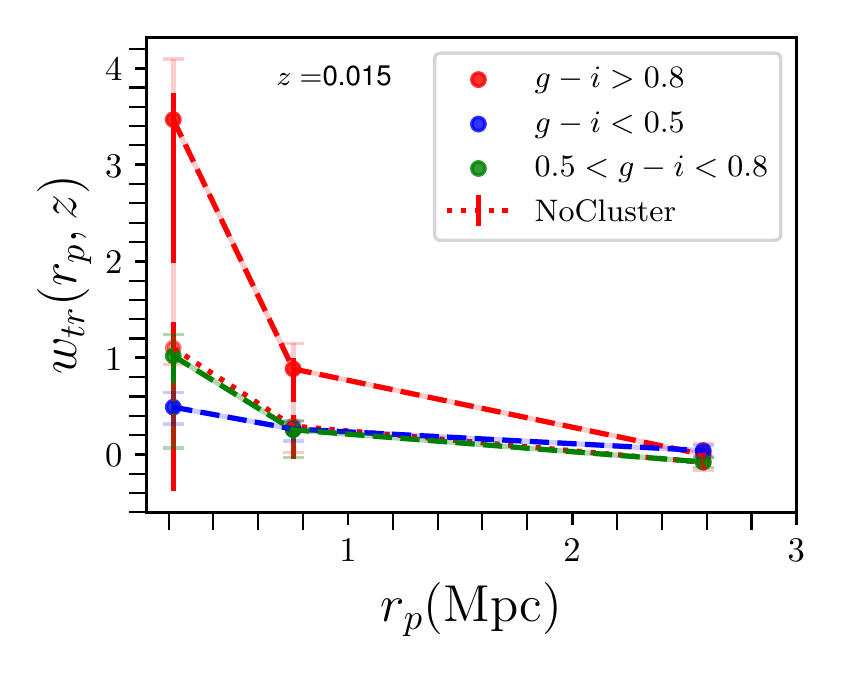}
    \includegraphics[width=0.45\linewidth]{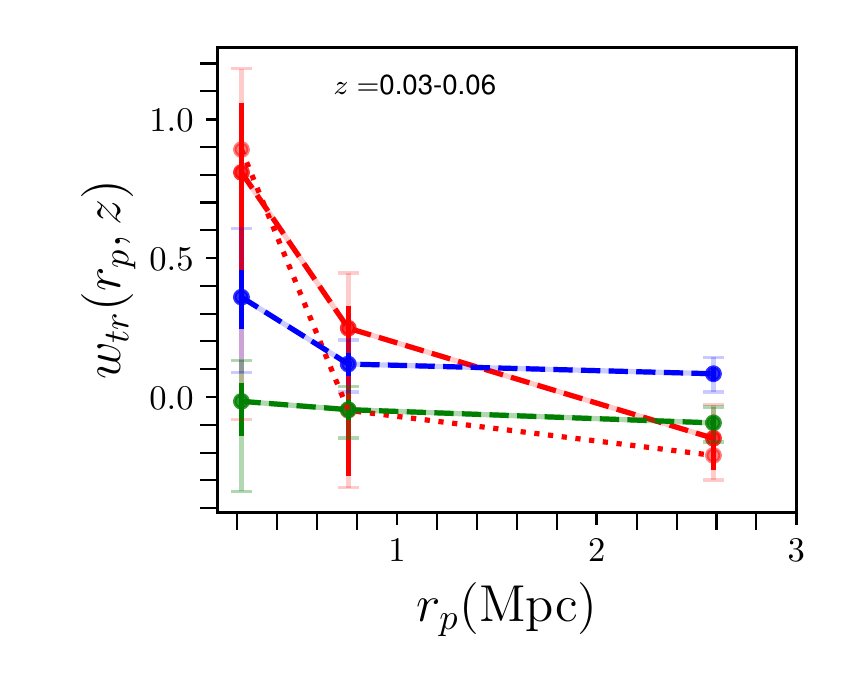}
\end{minipage}
\begin{minipage}{.98\textwidth}
    \centering
    \includegraphics[width=0.45\linewidth]{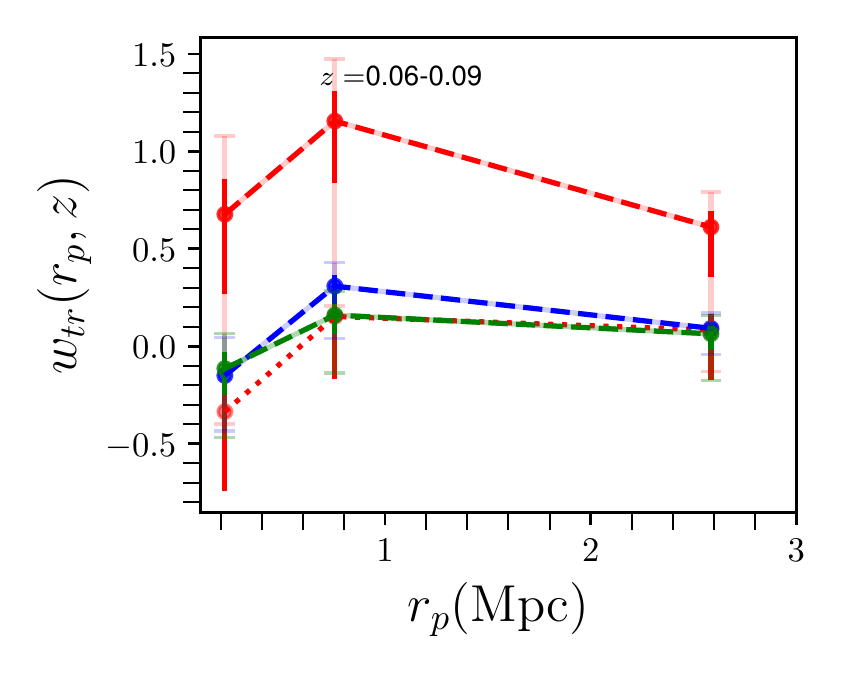}
    \includegraphics[width=0.45\linewidth]{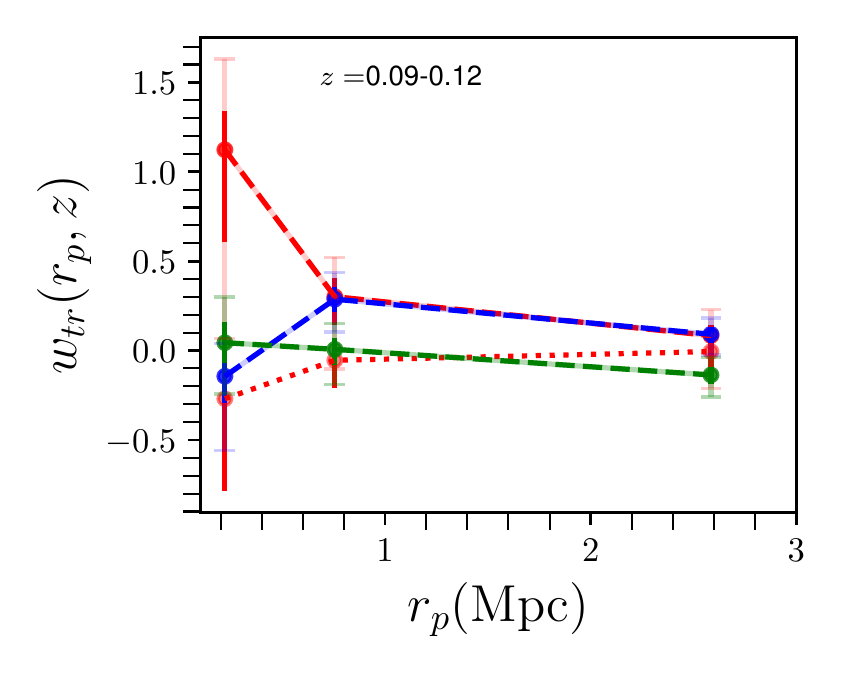}
\end{minipage}
\begin{minipage}{.98\textwidth}
    \hskip -1.0cm
    \centering
    \includegraphics[width=0.5\linewidth]{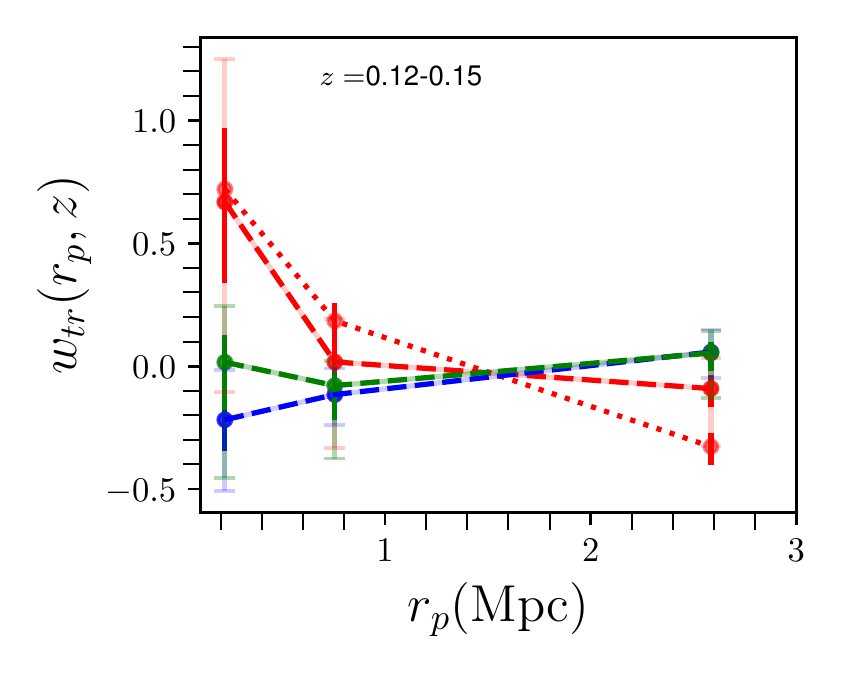}
\end{minipage}
\caption{Cross-correlation between LSB galaxies and the NSA catalog taken in five redshift bins (as indicated), measured separately for red ($g-i \geq 0.8$), blue ($g-i<0.5$), and green ($0.5 \leq g-i<0.8$) galaxies. We show $1 \, \sigma$ errors in bold, and $2 \, \sigma$ errors in semi-transparent color. We see a clear clustering signal at $z<0.03$ in all galaxy color ranges. The clustering strength rapidly drops thereafter for the green and then the blue sample. There is an additional peak at $z \sim 0.1$, particularly for the red sample. We also show the cross-correlation for red galaxies after removing LSB galaxies that are close to clusters in projection (\S \ref{sec:cluster}).
\label{fig:xcorcolor}}
\end{figure*}
%%%%%%%%%%%%%%%%%%%%%%%%%%%%%%%%%%

\subsection{Cross-correlation redshifts}

If an unknown sample is physically co-located with a tracer sample of known redshift, then we detect an angular correlation signal between them, while if they are at disjoint distances, they will show no excess clustering over a random sample of positions. This approach has been in use for a long time, including to understand the nature of rare objects like quasars or sub-millimeter galaxies \citep[e.g.,][]{Seldner:1979aa,Ho:2008aa,Mitchell-Wynne:2012aa}, to calibrate photometric redshifts \citep{Schneider:2006aa,Quadri:2010aa,Matthews:2010aa}, to determine redshift distributions for classes of galaxies \citep[e.g.,][]{Newman:2008aa} or to measure galaxy mass functions \citep{Bates:2019aa}.

We will follow the treatment outlined in \citet{Menard:2013aa} and \citet{Schmidt:2013aa}. They focus on the scenario that the target population (in our case LSB galaxies) span a relatively narrow redshift range, such that the evolution in galaxy bias of both the tracer and unknown populations can be ignored. They show that the clustering in the non-linear regime can also be useful for this application, and advocate examining the cross-correlation on scales larger than the survey point-spread function and smaller than the typical distance between reference objects, such that issues like Galactic reddening, cirrus or photometric zeropoints do not become serious systematics as they would on scales of $> 1$ degree. 

Taking our LSB galaxy sample as the ``target'' sample, with number density $n_t$, and the NSA sample as the ``reference'' sample with number density $n_r$, the correlation function of angle $\theta$ and redshift $z_i$ is:

\begin{equation}\label{eq:w}
w_{tr}(\theta,z_i) = \frac{\langle n_t (\theta, z_i) \rangle}{n_r} - 1
\end{equation}

\noindent
where the $\langle n_t (\theta, z_i) \rangle$ represents the average density of target galaxies around reference galaxies at redshift $z_i$.
The $w_{tr}(\theta,z_i)$ are estimated from pair counts between the target and reference samples, and the random samples that define the geometry of each survey (see \S \ref{sec:Gamatest}). Once we determine $w_{tr} (\theta, z_i)$, we will approximate the average cluster-redshift signal in a given redshift bin as:

\begin{equation}\label{eq:wbar}
\bar{w}_{tr}(z_i) = \frac{\int_{\theta_{\rm min}}^{\theta_{\rm max}} W(\theta) w_{tr} (\theta,z_i) \mathrm{d}\theta}{\int_{\theta_{\rm min}}^{\theta_{\rm max}} W(\theta) \mathrm{d}\theta}
\end{equation}

with $W(\theta)$ a filter that we apply to maximize the signal to noise, and we adopt a power-law to match the common form for galaxies. In reality, the cluster-redshift signal is an integral over the product of the bias in each population with the dark matter autocorrelation function. However, \citet{Menard:2013aa} argue that so long as the number density of the source varies more rapidly than evolution in the bias, then we can ignore bias terms and approximate the redshift distribution directly from the clustering amplitudes as a function of redshift.

To measure the clustering amplitudes, we turn to our primary tracer sample, the NSA. We measure the clustering signal using the cross-correlation estimator introduced by \citet{Davis:1983aa}:

\begin{equation}
\label{eq:davis}
w_{tr}(\theta_i, z_i) \approx \frac{D_r D_t}{D_r R_t} \frac{N_{R_t}}{N_t} -1.
\end{equation}

Here $D_r D_t$ represents the number of reference-target pairs separated by angular distance $\theta$, normalized by the total number of galaxies in each sample and $D_r R_t$ represents the number of pairs between the tracer and random positions in the target population search area, normalized by the number of each. This estimator does not account for the spatial selection function of the reference sample, since we do not know the NSA selection function. 

We also use the SDSS DR12 large-scale structure catalog. Here, we do understand the spatial distribution of the tracer SDSS sample, and so we use the \citet{Landy:1993aa} estimator:
\begin{multline}
  w_{tr}(\theta, z_i) \approx \frac{1}{R_r R_t} \\
  \left[ D_r D_t  \left( \frac{{N_R}_r {N_R}_t}{N_r N_t} \right) - D_r R_t \left( \frac{{N_r}_r}{N_r} \right) - D_t R_r \left( \frac{{N_R}_t}{N_t} \right) R_r R_t \right].
\end{multline}

The $DD$ and $DR$ terms again represent the number of target-reference and target-random pairs that are separated by a distance $\theta$ within a redshift slice $z_i$. In this case, we have random fields for both the reference and target samples. The cross-correlations ensuing from the SDSS DR12 cross-correlation are shown in Appendix A.

We integrate over a uniform projected physical range in each bin. As an inner angular scale, we take $\theta_{\rm min} = 0.10$~Mpc (or 1 arcmin at $z = 0.15$). Because of how we mask luminous galaxies prior to detection, this is the smallest angular scale over which we can hope to find a nearby LSB. As an outer scale, we integrate out to 3 Mpc. In part, this radius is chosen because we do not in general detect signal at larger scales, but is also motivated by the findings of \citet{Schmidt:2013aa}. On larger scales, issues like cirrus and photometric calibration can become an issue, so even though more real pairs are added, the chance of increasing systematic bias and degrading the overall signal-to-noise goes up. 

We calculate the redshift distribution as $\mathrm{d}N/\mathrm{d}z \propto \bar{w}_{tr} (z_i)$, where $\bar{w}_{tr}$ is the average clustering signal in a given redshift bin over the range of scales that we consider. We normalize the redshift distribution such that $\int dz \, dN/dz = 1$. We also note that the bias in both the reference and target samples are assumed to be constant over the full redshift range considered, which is probably a reasonable assumption out to $z \sim 0.15$. We verify that we can recover reasonable redshift distributions by building test samples with GAMA, where the distances are known, and then recovering them through cross-correlation (see Appendix A). These tests give us some confidence that even with small and very low redshift samples, it is possible to recover redshift distributions for faint galaxies.

%%%%%%%%%%%%%%%%%%%%%%%%%%%%%%%%%%
\begin{figure*}
\includegraphics[width=0.35\textwidth]{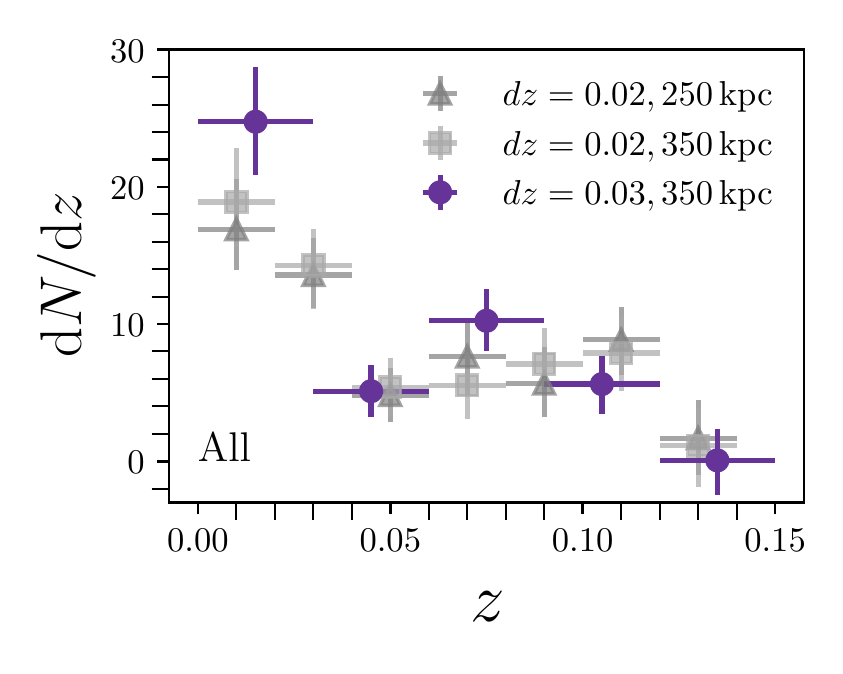}
\includegraphics[width=0.35\textwidth]{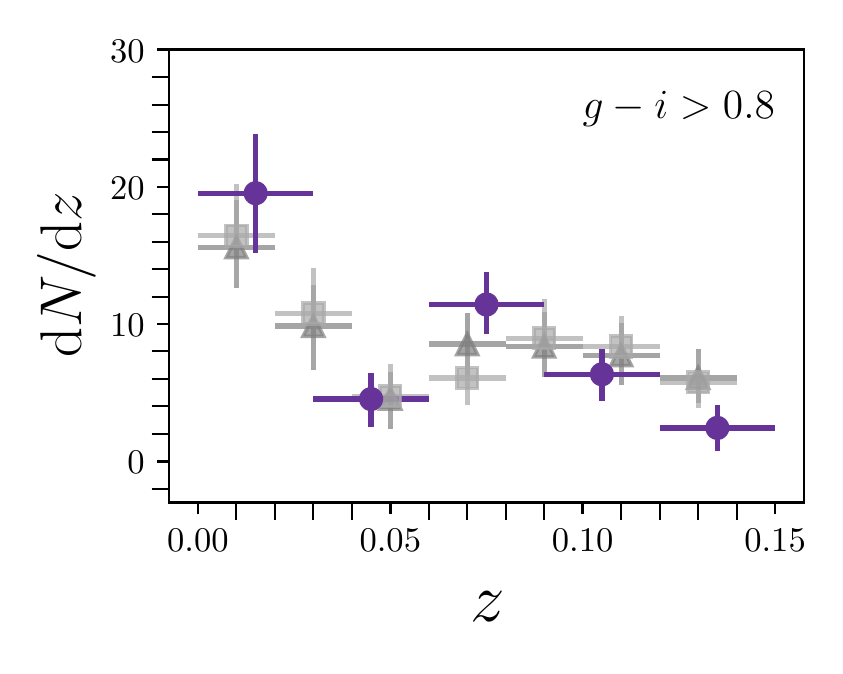}
\includegraphics[width=0.35\textwidth]{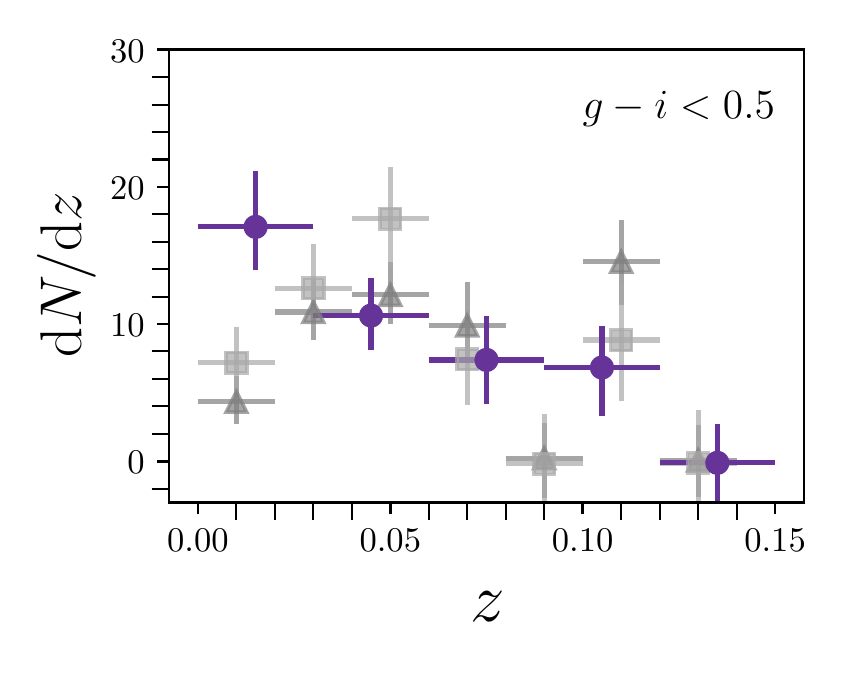}
\caption{Our inferred redshift distribution for the full sample including all (left), red (middle), and blue (right) galaxies. We compare the results for three different binning prescriptions (in both velocity and spatial binning), using $\delta z = 0.02, 0.03$ and inner bin $250, 350$ kpc. Error bars represent the 68\% range incorporating uncertainty in the $w_{tr}$ values. Following Equation \ref{eq:wbar}, we calculate the average clustering signal in each redshift bin. In the case of the red galaxies, the overall redshift distribution is stable to different binning decisions, but this is not the case for the blue galaxies, where the peak distance moves a bit under different binning prescriptions. In practice, our primary conclusions do not change if we carry through the different redshift distributions. 
\label{fig:wbar}}
\end{figure*}
%%%%%%%%%%%%%%%%%%%%%%%%%%%%%%%%%%

\subsection{LSB Cross-correlation with NSA}

As described in \S \ref{sec:sample}, we do not have a detailed mask for the NSA reference sample, and so use the Davis \& Peebles estimator (Eq.\ \ref{eq:davis}). We do not include all 781 LSBGs in this cross-correlation, but rather use the completeness map shown in \S \ref{sec:complete} to select a sample with completeness $>70\%$, which contains 339 galaxies. To calculate the cross-correlation uncertainties, we utilize the jacknife resampling procedure in the {\tt dsigma} galaxy-galaxy lensing tool\footnote{\url{https://github.com/johannesulf/dsigma}}, which computes contiguous regions, and then allows us to divide the total area into 100 sub-regions. We then recalculate the clustering 1000 times, removing 10 of the fields each time, to derive the errors. 

The clustering amplitude differs between red and blue LSB galaxies as seen in normal galaxies \citep[e.g.,][]{Zehavi:2005aa} as well as complementary observations of LSB galaxies \citep{Zaritsky:2019aa,Tanoglidis:2021aa}. Therefore, we calculate the clustering as a function of galaxy color. Galaxy color is measured from the matched \sersic{} fits. As described in G18, there are a number of clear differences between the red and blue samples. The blue galaxies are clearly forming stars, as indicated by detected H$\alpha$ in a handful of cases \citep{Greco:2018ab} or by the high \emph{GALEX} detection fraction among the blue galaxies. The red galaxies also extend to far lower surface brightness ($\mu_{\rm eff, g} < 28, 25.5$~mag~arcsec$^{-2}$ for red, blue respectively). While we see a clear bimodality in the color distributions in G18, there are significant errors on the color measurements and some inevitable mixing in color. We therefore focus on the extreme color ranges where even photometric errors cannot change the galaxy color. We divide the sample into three color ranges: 59 red ($g-i \geq 0.8$), 145 blue ($g-i<0.5$), and 135 green ($0.5\leq g-i<0.8$) galaxies. We calculate the cross-correlation in each redshift bin for each galaxy sample separately (Fig.\ \ref{fig:xcorcolor}). The clustering strength for the red galaxies is $\sim 6-7$ times higher than for the blue galaxies at $z<0.03$.  

It is not surprising to find higher clustering among the red galaxies. As shown by \citet{Geha:2012aa}, low-mass central galaxies are nearly always star-forming, while low-mass satellites in groups and clusters tend to be quenched and therefore red. The blue galaxies live in lower-density environments and have a lower galaxy bias, and thus the clustering signal is weaker and more challenging to detect. 

We only detect a significant clustering signal for the green galaxies for $z<0.03$ ($w_{tr} = 1.0 \pm 0.1$).  These galaxies, which live between the red and blue galaxies in color space, show less clustering signal than either the red or the blue galaxies. We do not have a full explanation for their lack of clustering in this redshift range, but discuss some possible factors here. 

If the green galaxies were populated fully by scatter from red and blue galaxies, we may expect that the differing bias and redshift distribution between the two samples might make the signal very challenging to detect. Another factor at play is that background spiral contamination may be most prevalent in the green galaxies, which could be intrinsically blue galaxies at slightly higher redshift. If we take the surface brightness distribution of spiral galaxies as measured by \citet{Macarthur:2003aa}, then the lower end of the distribution is $\mu_{\rm eff, B} \approx 23$~mag~arcsec$^{-2}$. Such a relatively low surface brightness spiral would be scattered into our sample through surface brightness dimming at $z>0.35$. At the same time, as the 4000\AA\ break moves into the $r-$band, the color will redden. This possibility is somewhat supported by the cross-correlation signal between the green galaxies and the SDSS DR12 LSS catalog, where we do see hints of a signal in the $z=0.32-0.34$ bins, perhaps indicating that there is a population of higher-redshift LSB spirals that has crept into our sample.

We attempt to construct a cleaner sample, in which we visually flag likely contaminants (see Appendix B for examples). These include both galaxies that appear bulge-dominated and a handful of objects that are likely to be tidal debris \citep{Greco:2018ab}. We find no difference in the cross-correlation signal for the sample with likely contaminants masked.

A final possibility is that the green galaxies are physically distinct from the red and blue ones. They may be more likely to be ``back-splash'' galaxies \citep{Benavides:2021aa}, that have tidally interacted with a more massive galaxy but are not technically within its virial radius, or even perhaps beyond the $\sim 1$ Mpc scales that dominate our clustering signal. Larger samples will be needed to determine the nature of these green galaxies, and we will focus more strongly on the red and blue subsets for the remainder of the paper.

We also confirm the clustering results using the SDSS DR12 Large Scale Structure sample, that contains a detailed spatial mask. Since the area of overlap is smaller by $\sim 25\%$, we only use these as a sanity check, but they do confirm the trends presented here (see Appendix B for details.)

\subsection{The redshift distribution}

We now infer the redshift distribution directly from the average clustering strength in each bin as $\mathrm{d}N/\mathrm{d}z \propto \bar{w_{tr}}$. To calculate $\bar{w_{tr}}$, we use Eq. \ref{eq:wbar}, and as our matched filter we take $W(\theta) \propto \theta^{-0.8}$, since this is often a good fit to galaxy correlation functions \cite[e.g.,][]{Peebles:1974aa,Zehavi:2005aa}. In Figure \ref{fig:wbar}, we plot the average cross-correlation signal for the full, red, and blue samples, normalized such that $\int dz (dN/dz) = 1$. The errors represent the 68\% distribution incorporating uncertainties in the $w_{tr}$ measurements.

The redshift distribution peaks at $z<0.02$ ($D=100$~Mpc). For the red galaxies, we find that $\sim 30\%$ of the galaxies have $z<0.02$, $\sim 20\%$ have $0.02<z<0.05$, and $\sim 40\%$ have $z>0.05$, with nominal uncertainties of $\sim 15\%$. The redshift distribution appears flatter for the blue galaxies, with 20-40\% of the sample falling in each of these three redshift bins. Looking at the $\delta z = 0.02, 350~\rm{kpc}$ bins, the inferred redshift distribution is marginally different, 10-25\% of the sources falling in the first redshift bin, 30-60\% in the second redshift bin, and 20-50\% in the third redshift bin. This distribution is still consistent at $1 \, \sigma$ with the default distributions. Within errors, the distributions are consistent with each other regardless of binning. The NSA does not extend beyond $z = 0.15$, but in Appendix C we show that there is minimal correlation signal beyond $z \sim 0.15$, and thus by happenstance the NSA does provide optimal distance coverage for our purposes. 

Galaxies within $D<100$~Mpc are typically ``normal'' $\sim 10^7$~\msun\ galaxies, in the sense that their size is similar to known galaxies in this mass range \citep[e.g.,][C21 hereafter]{Carlsten:2021aa}, whose intrinsically low surface brightnesses place them within our selection. Galaxies at larger distance, in contrast, are typically more massive and are outliers from the median mass-size relation. Ultra-diffuse galaxies, as defined by \citet{van-Dokkum:2015aa} populate the higher-redshift tail of our distribution.

The sample of independently measured redshifts remains quite small at present, particularly at the lower surface brightness end of our sample. However, we have performed \emph{Hubble Space Telescope} follow-up of LSB dwarfs  from \citet{Greco:2018aa} that were close in projected distance to two $z \sim 0.08$ groups. \citet{Somalwar:2020aa} found globular cluster populations consistent with these LSB galaxies likely being at the distance of their groups, providing some evidence for a higher redshift tail that extends at least to $z \sim 0.1$. In \S \ref{sec:cluster} we will investigate the LSB galaxies that are likely to be associated with rich groups and clusters, to again see evidence for a tail of galaxies extending to $z \approx 0.15$.

%%%%%%%%%%%%%%%%%%%%%%%%%%%%%%%%%%
\begin{figure*}
\begin{minipage}{.98\textwidth}
    \centering
    \includegraphics[width=0.45\linewidth]{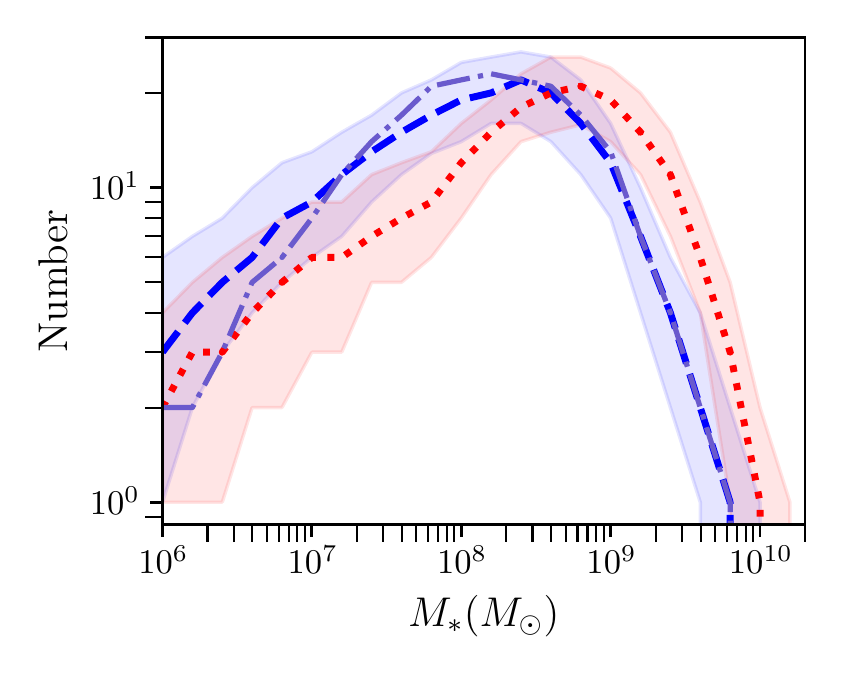}
    \includegraphics[width=0.45\linewidth]{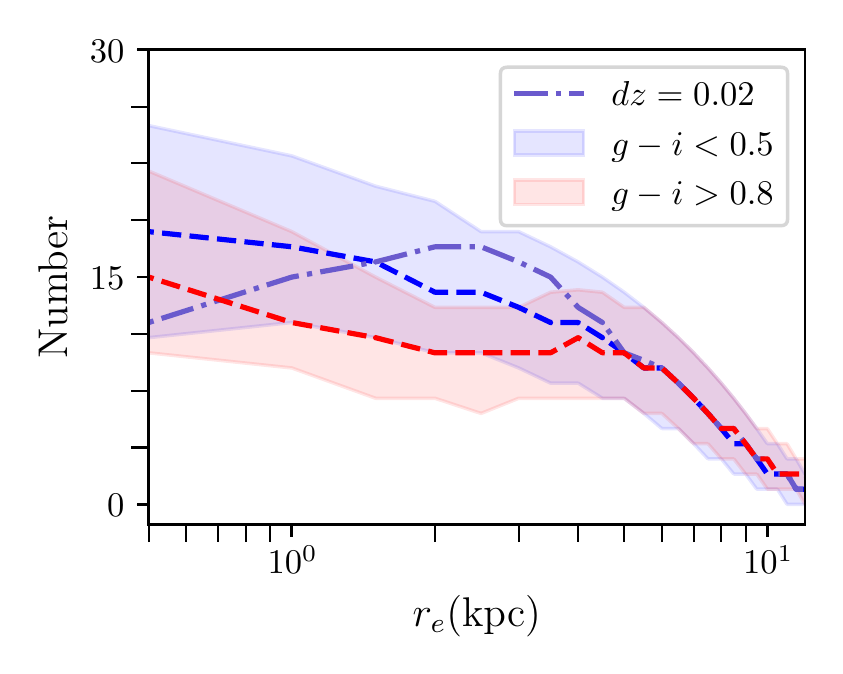}
\end{minipage}
\begin{minipage}{.98\textwidth}
    \centering
    \includegraphics[width=0.46\linewidth]{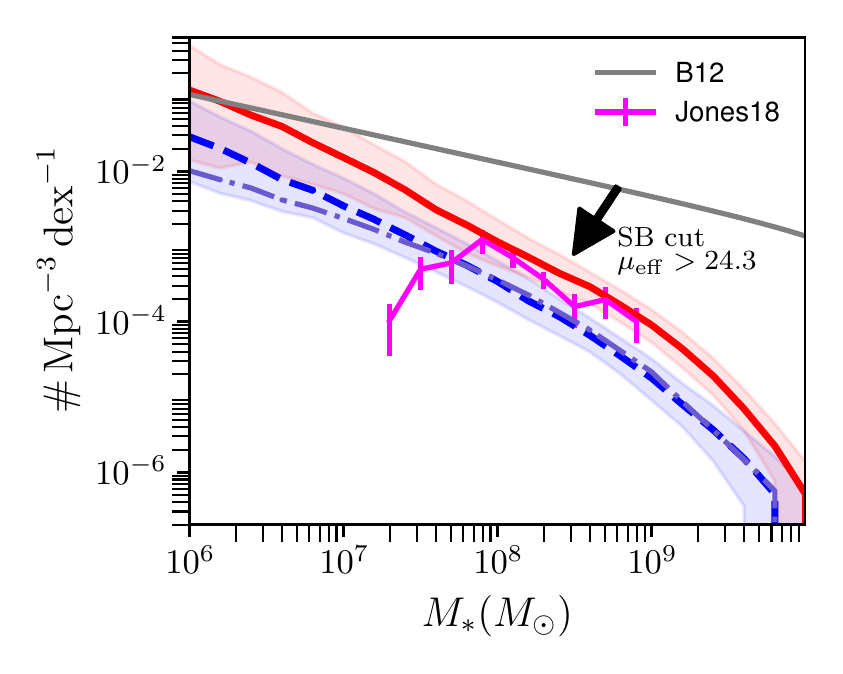}
    \includegraphics[width=0.46\linewidth]{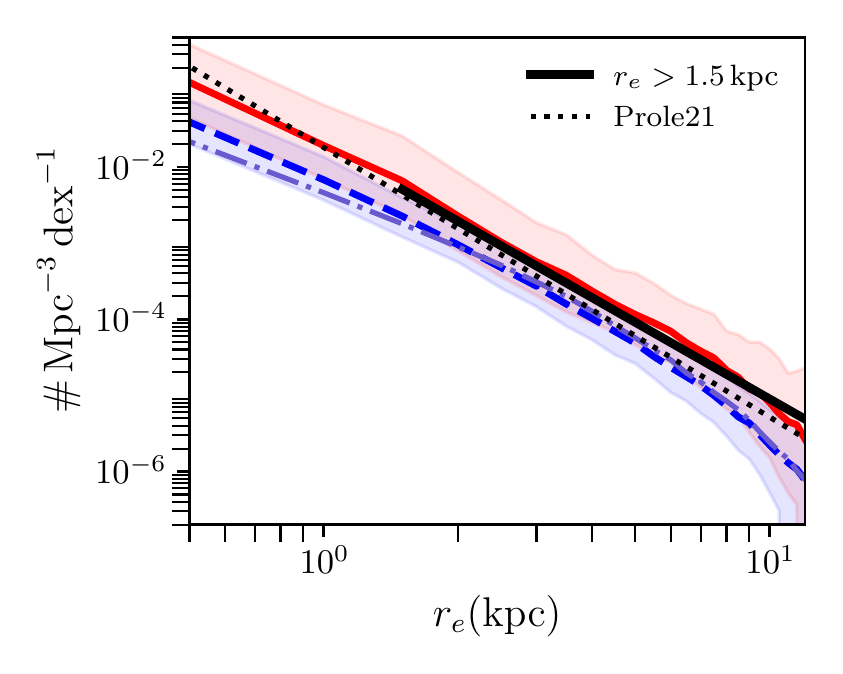}
\end{minipage}
\caption{Inferred distribution in stellar mass and size given the redshift distributions from the cross-correlation, based on $d z = 0.03$ and three spatial bins. To show that the precise redshift binning does not strongly impact our results, in dot-dash, we show the difference in the inferred distributions for the blue galaxies if we use instead the redshift distributions based on $d z = 0.02$ bins. On the top row we show the inferred observed distributions. On the bottom row, we show the volume and completeness-weighted distributions. On the bottom left, we compare with the stellar mass function inferred for an \ion{H}{1} selected sample of UDGs \citep[][]{Leisman:2017aa,Jones:2018aa}. On the bottom right, we compare with the slope fitted to a complementary LSB sample selected by \citet{Prole:2021ab} from HSC.
\label{fig:mass_size_hist}}
\end{figure*}
%%%%%%%%%%%%%%%%%%%%%%%%%%%%%%%%%%

\section{Physical Properties of the LSB Galaxy Sample}
\label{sec:physicalprop}

We have used cross-correlation with a known tracer population to calculate the redshift distribution for the ensemble of LSB galaxies discovered in \citet{Greco:2018aa}. In this section, we combine the redshift distribution with the observed angular sizes, magnitudes, and colors of each galaxy to infer their mass and size distributions. As argued above, it is more meaningful to treat the red and blue galaxies separately, given their differing bias and possibly different redshift distribution. In this analysis, we utilize 762 of the 781 galaxies, excluding a small number of the lowest surface brightness sources with problematic photometry and inferred completeness levels $c<0.01$.

\subsection{Mass and Size}

%%%%%%%%%%%%%%%%%%%%%%%%%%%%%%%%%%
\begin{figure*}
\includegraphics[width=0.9\textwidth]{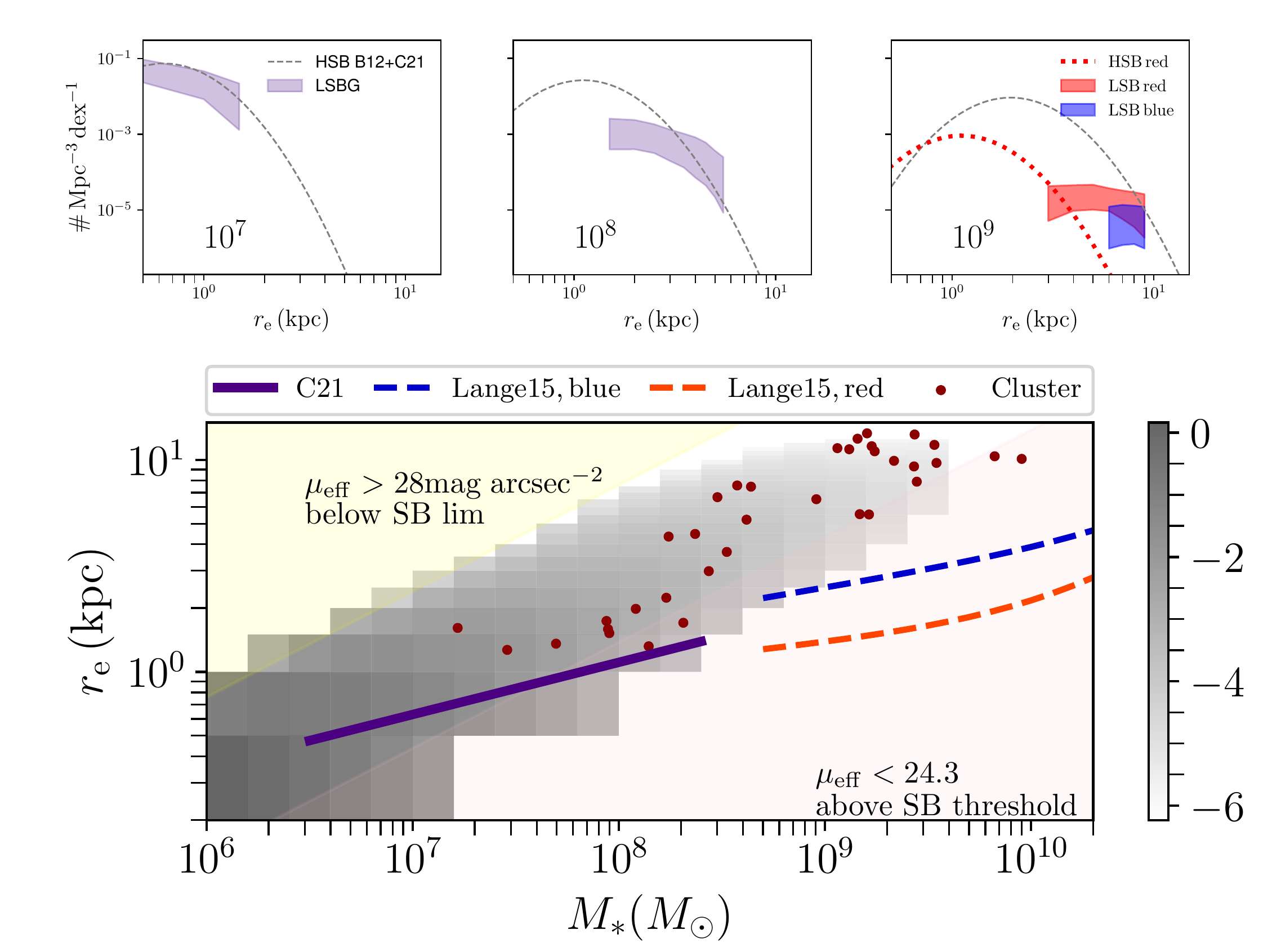}
\caption{{\it Bottom}: Distribution in the mass-size plane for the full sample of LSB galaxies. The color scale represents the volume weight for each mass-size bin. Points are red ($g-i>0.8$) galaxies that are within a projected 0.7 Mpc from groups within $z<0.15$, presuming each is at the distance of their cluster. We show the C21 (purple) and \citet{Lange:2015aa} (red/blue) mass-size relations for comparison, and indicate tracks of constant surface brightness corresponding roughly to our selection region at a typical distance of 200 Mpc (white lines). Light green regions are excluded from our selection for either being below our surface brightness sensitivity (upper left) or too high in surface brightness (bottom right). Grey-scale represents the number density in that cell.
{\it Top}: We plot slices (in dashed lines) through the mass-size relation at $10^7, \, 10^8, \, 10^9$~\msun\ using the 0.2 dex log-normal scatter from C21, and normalizing the overall number density to \citet{Baldry:2012aa}. Above $M_*=10^9$~\msun\ we assume that red galaxies account for 10\% of the population. Shaded regions indicate the range of number densities possible for LSB galaxies in the same mass range. We see an excess of red LSB galaxies over expectations based on mass-size fits to known galaxy populations.
} 
\label{fig:mass-size}
\end{figure*}
%%%%%%%%%%%%%%%%%%%%%%%%%%%%%%%%%%

We first present the distributions in mass and size separately, and then examine their joint distribution. To construct the mass and size histograms, we follow C21 and use the color to assign a stellar mass following \citet{Into:2013aa}. It is hard to get a robust estimate of the scatter introduced by this relation, since color-mass relations are calibrated for more massive systems. 
In a forthcoming work, we directly compute the color-mass relation of a set of \ion{H}{1}-selected UDGs and NSA dwarfs (Kado-Fong et al., 2022; in preparation). We compute the stellar mass of each galaxy via UV-optical SED fitting assuming a \cite{kroupa:2001} IMF using \textit{Galex} and HSC-SSP imaging in conjunction with the \ion{H}{1}-derived distances. We find no significant difference between the color-mass relation of the UDG and spectroscopic dwarf samples. Our measured color-mass relation lies between the \cite{Into:2013aa} and \cite{Bell:2003aa} color-mass relations (the former predicts lower mass-to-light ratios at fixed $g-r$ color), and falls within 2$\sigma$ of both relations. In particular, the deviation from the \cite{Into:2013aa} relation for $\Delta \Psi=\Delta \log_{10}(M_\star/L_g)$ has a mean value of $\langle \Delta \Psi \rangle= 0.10(0.08)$ dex for the UDG (spectroscopic) sample with a standard deviation of $\sigma_{\Delta \Psi} = 0.08(0.12)$ dex. Given that the 4000\AA\ break is within the $g-$band over our full distance range, and given our ignorance of the galaxy spectra, we ignore $k$-correction and treat the observed colors as the rest-frame values. 
 
To construct the mass and size distributions, we first create an interpolated redshift distribution. We then randomly assign a distance to each galaxy drawn from that distribution over a thousand draws. We incorporate scatter in color and size from G18. We generate histograms (unweighted), and then we correct each galaxy for surface brightness incompleteness (Figure \ref{fig:complete}) and weight by the maximum observable volume as described in \S \ref{sec:complete} (Figure \ref{fig:mass_size_hist}). For completeness, we also carry through these calculations using the $\rm{d}z = 0.02$ binning for the blue galaxies, since the redshift distributions are marginaly different, to show that the final results are insensitive to our binning choices.

We had some concern that we would end up with unphysically large galaxies, given the long tail in our redshift distribution. Thus, we also experimented with ranking the galaxies by apparent size, and enforcing an upper size limit (and thus upper distance limit) for each galaxy. Almost none of the conclusions presented here change with that constraint, but the slope in the size distribution does become steeper than that observed by other groups. If there is an upper size limit, it is not known, and since $80 \pm 4 \%$ ($90 \pm 3 \%$) of our galaxies are smaller than 8 (10) kpc, we do not impose this additional constraint. 

To examine the maximum galaxy size with a bit more care, we assume that our inferred redshift distribution is correct, and that there are G18 galaxies extending to $z \approx 0.15$. We look at the inferred number of galaxies with $7<r_{\rm e}/\rm{kpc}<8$ and $8<r_{\rm e}/\rm{kpc}<9$ accounting for both the uncertainty in the redshift distribution and in the size measurements, which are typically $\sim 1 \arcsec$ \citep{Greco:2018aa}. We focus here only on the red galaxies because their clustering signal is much more robustly measured. Also, it is very hard to imagine a background contaminant to these galaxies, since such a population would constitute physically huge and red galaxies. For the red subsample, we find 10$\pm 3$ galaxies with $7<r_{\rm e}/\rm{kpc}<8$, a $3 \, \sigma$ detection. We conservatively do not apply a completeness correction to derive this number. We find $8 \pm 3$ galaxies in the next higher bin, so we do not have a significant detection for $r_{\rm e} > 8$~kpc. Therefore, the maximum galaxy size in this sample is $\approx 8$~kpc. This is an extreme galaxy size, even larger than most UDGs in Coma. On the other hand, these galaxies are rare, with number densities $n < 10^{-5}$~Mpc$^{-3}$.

We also briefly investigate whether a systematic error in our sizes would shift the size distribution dramatically, by shifting all the sizes downward by $\sim 0.5$\arcsec, or half the size uncertainty. In this case, we find $9 \pm 3$ galaxies with $7<R_e/\rm{kpc}<8$, maintaining the $3 \, \sigma$ detection. 

We thus infer a stellar mass range of $10^7-10^9$~\msun\ and sizes of $r_e \approx 1-8$~kpc. This mass range is comparable to other LSB dwarf galaxy samples where distances, masses, and sizes are inferred from clusters \citep[e.g.,][]{van-Dokkum:2015aa,van-der-Burg:2016aa,Koda:2015aa} or distances are measured from \ion{H}{1} \citep[e.g.,][]{Leisman:2017aa,Karunakaran:2020aa,Kado-Fong:2021aa} or optical spectroscopic follow-up \citep[][]{Kadowaki:2017aa,Kadowaki:2021aa}. A substantial fraction of our sources do technically qualify as ultra-diffuse galaxies based on our surface brightness limit of $\mu_{\rm g,eff} > 24.3$~mag~arcsec$^{-2}$ and inferred size $r_e>1.5$~kpc. Specifically, $65-80\%$ of the red galaxies, and $60-77\%$ of the blue galaxies fall above the $1.5$~kpc size boundary by number. 

The differing distributions in redshift, angular size, magnitude, and color between the red and blue samples translate into small differences in the mass distributions for the two samples. Specifically, the peak mass of the red galaxies is shifted to slightly higher mass due to the second higher-redshift peak in the inferred redshift distribution. There is a corresponding very small shift to larger size for the red galaxies as well.

We compare our inferred stellar mass function with two literature mass functions (Figure \ref{fig:mass_size_hist}, bottom left). First, we compare to the galaxy mass function from \citet{Baldry:2012aa} \citep[see also][]{Wright:2017aa}. We see that at $M_* \sim 10^7$~\msun, our selection finds the same number density of galaxies as are inferred to be in the field. However, as we go to higher mass, our surface brightness selection increasingly selects rarer subsets of galaxies. We also compare to the mass function from \citet{Jones:2018aa} for \ion{H}{1} selected UDGs from \citet{Leisman:2017aa}. We find reasonable agreement with their mass function above $M_* \approx 10^8$~\msun, which is rather remarkable given the differences in how the stellar masses are calculated. In any case, we come to a similar conclusion to Jones et al.\ in that the LSB galaxies make up a small fraction of the galaxy population at this mass.

The LSB size distribution is well-described by a power-law. Fitting only for galaxies with $r_{\rm eff} > 1$~kpc, we fit a slope of $\alpha = -3.3 \pm 0.6$ for $N \propto r_{\rm e}^{\alpha}$, consistent with the slope measured by \citet[][$-3.5 \pm 0.3$]{Prole:2021aa} to their own HSC-selected LSB sample \citep{Prole:2021ab}. A similar slope was reported from the cluster sample of \citet{van-der-Burg:2017aa} and \citet{ManceraPina:2019aa}, while \citet{Amorisco:2016aa} predict a similar slope if UDGs form in halos with the highest spin. It is worth noting that our slope is sensitive to the exact size range that we fit over, and in particular the slope is shallower at smaller size.

We then combine the mass and size constraints in Figure \ref{fig:mass-size}. Completeness corrections and volume weights are applied. First of all, we see that the slope of the mass-size distribution for this sample of LSB galaxies appears to be steeper than the measured slope from C21 or \citet{Lange:2015aa}. This slope is not an attribute of the sample, but rather a direct artifact of our surface brightness selection. To make this point, we indicate excluded regions in surface brightness for galaxies at 200 Mpc. Technically, all cells above the $\mu_e = 24$ and $r_e > 1.5$~kpc lines qualify as UDGs as defined by \citet{van-Dokkum:2015aa}. Nearly all the galaxies with $M_* > 10^8$~\msun\ are in this category. This tail of more massive and more extended galaxies is a direct consequence of the cross-correlation signal that we detect at $z>0.05$. Most known galaxies with $M_* < 10^{7.5}$~\msun\ fall within our selection, so we are not selecting outliers in that mass range, at least according to the mass-size relations of C21. At higher mass, in contrast, we are mostly sensitive to galaxies that are considerably larger than the ridge-line of the mass-size relation. 

Direct comparisons with typical galaxies are made more difficult by the lack of a robust measurement of the mass-size relation across all environments and the full mass range probed by our sample \citep[for instance,][are both incomplete below $\sim 10^9$~\msun]{Shen:2003aa, Lange:2015aa}. While C21 showed that there is little difference in the mass-size relation between early and late-type galaxies for masses $10^{5.5}-10^{8.5}$~\msun\ \citep[see also][]{Kormedy:2016aa}, the situation changes dramatically at higher mass. Early-type galaxies flatten to a constant size of $\sim 1$~kpc between $\sim 10^{8.5}-10^{10}$~\msun. This flattening has been seen in large environment-blind volumes \citep[e.g.,][]{Shen:2003aa}, in individual cluster galaxies \citep[e.g.,][]{SmithCastelli:2008aa,Misgeld:2008aa,Eigenthaler:2018aa}, and even out to intermediate redshift \citep[][]{Nedkova:2021aa}. On the other hand, late-type galaxies continue to grow in size over this $10^{8.5}-10^{10}$~\msun\ mass range. The fits from \citet[][red and blue dashed lines in Figure \ref{fig:mass-size}]{Lange:2015aa} show that between the high-mass end of C21 and $\sim 5 \times 10^9$~\msun, galaxies are $\sim 1$~kpc in size. We now turn to examine the size distributions at fixed mass.

%%%%%%%%%%%%%%%%%%%%%%%%%%%%%%%%%%%%%%%%%%%
\begin{deluxetable*}{cccc}
\tablenum{7}
\tablecaption{LSB Galaxy Mass Function\label{tab:redmass}}
\tablewidth{0pt}
\tablehead{
  \colhead{log Mass} &\colhead{ \# (low)} & \colhead{\# (med)} & \colhead{\# (high)} \\
\colhead{($M_{\odot}$)} & \colhead{($10^{-3}$ Mpc$^{-2}$ dex$^{-1}$)} & \colhead{($10^{-3}$ Mpc$^{-2}$ dex$^{-1}$)} & \colhead{($10^{-3}$ Mpc$^{-2}$ dex$^{-1}$)} 
}
\decimalcolnumbers
\startdata
7.00 & 5.032 & 15.203  & 38.669 \\
7.20 & 3.171 & 9.573  & 22.373 \\
7.40 & 2.409 & 5.662  & 13.401 \\
7.60 & 1.387 & 3.081  & 6.727 \\
7.80 & 0.813 & 1.931  & 4.037 \\
8.00 & 0.551 & 1.154  & 2.264 \\
8.20 & 0.389 & 0.721  & 1.288 \\
8.40 & 0.251 & 0.442  & 0.781 \\
8.60 & 0.174 & 0.292  & 0.454 \\
8.80 & 0.098 & 0.165  & 0.261 \\
9.00 & 0.053 & 0.090  & 0.145 \\
9.20 & 0.024 & 0.044  & 0.072 \\
9.40 & 0.011 & 0.019  & 0.033 \\
\hline
\hline
7.00 & 1.516 & 3.464  & 8.090 \\
7.20 & 1.080 & 2.290  & 5.000 \\
7.40 & 0.715 & 1.427  & 2.867 \\
7.60 & 0.459 & 0.867  & 1.776 \\
7.80 & 0.293 & 0.563  & 1.089 \\
8.00 & 0.181 & 0.343  & 0.643 \\
8.20 & 0.107 & 0.189  & 0.374 \\
8.40 & 0.065 & 0.117  & 0.202 \\
8.60 & 0.040 & 0.066  & 0.112 \\
8.80 & 0.020 & 0.035  & 0.060 \\
9.00 & 0.009 & 0.018  & 0.032 \\
9.20 & 0.004 & 0.008  & 0.015 \\
9.40 & 0.001 & 0.004  & 0.008 \\
\enddata
\tablecomments{Top is red mass function, bottom is blue. (1) Log of the stellar mass. (2) $1 \sigma$ low value of mass function. (3) Median mass function. (4) $1 \sigma$ high value of mass function.}
\end{deluxetable*}
%%%%%%%%%%%%%%%%%%%%%%%%%%%%%%%%%%%%%%%%%%%

\subsection{Size Distribution at Fixed Mass}

We do not yet know whether LSB galaxies are outliers in the mass-size relation \citep[e.g.,][]{van-der-Burg:2016aa,Danieli:2019aa}. In general, both in dwarfs and more massive galaxies, the size distribution at fixed mass appears to be well-fit by a log-normal \citep[e.g.,][C21]{Shen:2003aa}. We adopt the 0.2 dex scatter that was measured by C21, understanding that this value may not apply above $M_* \approx 10^{8.5}$~\msun\ \citep[although it is similar to the dispersion measured for more massive galaxies][]{Kawinwanichakij:2021aa}. We can then ask whether the LSB galaxy distribution is well-modeled as the tail of the log-normal that describes known galaxy samples. 

We therefore examine the size distribution of our sample galaxies in fixed mass {\bf slices}, and compare with ``normal'' galaxies (Figure \ref{fig:mass-size}). We plot the size distribution at fixed masses of $M_* = 10^7, 10^8, 10^9$~\msun\ for normal (dashed) and LSB (filled region) galaxies. The normal galaxies are assumed to have a log-normal size distribution, with the mean size and width given by C21. The normalization of the distribution is set such that the integral matches the number density at that mass from \citet{Baldry:2012aa}. The normalization of the LSB region is set by the mass functions presented in Figure \ref{fig:mass_size_hist}. Note that for $M_*<10^{8.5}$~\msun, C21 show that quenched/red and star-forming/blue galaxies have the same size distribution at fixed mass. We combine the red and blue samples for those mass bins. At the highest mass bin, red galaxies are $\sim 1$~kpc in size, while blue galaxies continue to grow larger with mass. Therefore, we compare those populations separately, assuming that red galaxies comprise $\sim 10\%$ of the galaxy population at $M_* \sim 10^9$~\msun\ \citep[e.g.,][]{Blanton:2009aa}. 

Starting with the left-most slice through the mass-size plane, we see that the $M_* = 10^7$~\msun\ LSB galaxies are comparable to the full galaxy sample. Both the number density and the size distributions match. We do not detect galaxies with low enough surface brightness to find any excess relative to the measured log-normal size distribution at this mass, should it exist. At $M_* = 10^8$~\msun\ there is a hint that the LSB galaxies may extend a bit beyond the normal galaxies, but better surface brightness sensitivity would still be needed to say for sure. Only at $M_* = 10^9$~\msun, where our sensitivity allows us to probe well beyond the peak in the mass-size relation, do we see clear evidence that the red galaxies are considerably larger than expected from the log normal size distribution. Normal red galaxies at $M_* \sim 10^{8.5}-10^{9.5}$~\msun\ have $r_{\rm e} \approx 1$~kpc in size, while our LSB galaxies range from $r_{\rm e} = 3-8$ kpc in this mass range (as argued above, the detection of galaxies with $r_{\rm e} > 8$~kpc is marginal). Given the observed number densities, this size range is totally inconsistent with a 0.2 dex log-normal size distribution centered at 1 kpc. 

It is hard to say whether something special is happening at $\sim 10^9$~\msun. On the one hand, transition happens at this mass. Below $\sim 10^9$~\msun, all galaxies have very similar mass-size relations. Above this mass, star-forming galaxies are becoming thin disks \citep{Kado-Fong:2020aa} while quenched galaxies have a flat mass-size relation. So perhaps, the excess of very extended LSB galaxies is also associated with this special mass range. On the other hand, there may be an extended tail of puffy galaxies at higher and lower mass that fall outside of the surface brightness sensitivity of our survey \citep[see, e.g.,][]{Danieli:2019aa}.

\subsection{A Likely Group and Cluster Sample}
\label{sec:cluster}

The G18 sample is ``blind'' to environmental density, which is one of its strengths. However, as discussed by \citet{Tanoglidis:2021aa}, cross-matching the full sample with known overdensities can also prove interesting. First, it provides a sanity check on derived physical properties, since we will show that the cluster sample is too numerous to have happened by chance. Second, we can ask what fraction of the red LSB galaxies are likely associated with groups and clusters. Third, we double check that the cross-correlation signal persists even with the cluster galaxies removed.

To identify groups and clusters, we use the \citet{Yang:2007aa} group and cluster catalog built from the seventh data release of the SDSS \citep{Abazajian:2009aa}. Halo masses in the catalog are calculated based on group richness and abundance matching. We select all halos with $M_h > 10^{13}$~\msun, and then identify all the LSB galaxies that lie within a projected 0.7 Mpc distance of these groups and clusters. We have chosen a fixed physical aperture for simplicity, and a relatively small size to limit contamination from chance alignments. To boost the likelihood that the galaxies are associated with clusters, we also restrict to the red LSB galaxies ($g-i>0.8$). There are 36 red galaxies matching these criteria. We find galaxies projected within clusters across the full redshift range to $z=0.15$. These galaxies encompass $\sim 16\%$ of the LSB galaxies in this color range, which is a lower limit to the fraction of red galaxies residing in groups and clusters, and similar to the 30\% within 1 Mpc of clusters reported by \citet{Tanoglidis:2021aa}. The projected cluster sample are plotted as points on the mass-size plane in Figure \ref{fig:mass-size} under the assumption that all of the satellites fall at the distance of their group. Their mass distribution skews a bit more massive than the overall sample, while the size distribution ranges from $1-10$~kpc  (Figure \ref{fig:mass-size}). Most of these galaxies would qualify as UDGs, should they indeed be at the distance of their associated group. 

We evaluate the probability to find this number of projections purely by chance. We perform the following Monte Carlo test 100 times. We select 36 random positions within our footprint, and match those positions with the same Yang catalog as the real galaxy positions. We find a mean of $17 \pm 4$ matches, making our detection of 36 matched galaxies a $4 \sigma$ outlier. We thus conclude that we are detecting a real sub-population of red UDGs in clusters. The fact that we find matches with clusters out to $z \sim 0.15$ may provide some additional confirmation of our inferred redshift range, although follow-up of those most distant cluster galaxy candidates would provide nice confirmation. 

We also recalculate the cross-correlation for the LSB sample excluding those close in projection to clusters (Figure \ref{fig:xcorcolor}). We still see a significant signal in the red galaxies ($w_{tr} = 1.1 \pm 0.1$ within 100 kpc and $z<0.015$). The clustering amplitude for the cluster-free sample is significant at $z<0.045$ and $z=0.12-0.15$. Thus, even excluding galaxies belonging to groups and clusters in projection, we recover a clustering signal for the red galaxies. 

\section{Discussion and Summary}

This paper aims to determine the distance distribution, via cross-correlation, of the LSB galaxies selected by \citet{Greco:2018aa} from all environments within the HSC-SSP Survey. We find a relatively broad distance distribution extending to $z \sim 0.15$ but peaked at $D<100$~Mpc. Spectroscopic follow-up for two G18 galaxies find $D<40$~Mpc, corresponding to the first prominent distance peak in our distribution. This distance distribution is consistent with inferences by \citet{Prole:2021ab} and \citet{Tanoglidis:2021aa}, based on similarly selected samples from HSC and DES respectively. It is broader than the distance distribution inferred for the Systematically Measuring Ultra-Diffuse Galaxies \citep[SMUDGes,][]{Zaritsky:2019aa,Zaritsky:2021aa} program, which does not seem to find very many galaxies more distant than 100 Mpc \citep{Barbosa:2020aa,Kadowaki:2021aa}. It is also broader than the distance distribution inferred for a 36 deg$^2$ area around the NGC 1052 group inferred by \citet{Roman:2021aa} from cross-correlation, but no spectroscopic tracers with $z>0.04$ were included in that work. From this distance range, we are also able to infer physical parameter distributions for the sample, including number densities, masses, and sizes.

We find typical stellar masses of $\sim 10^7-10^9$~\msun\ and a range of sizes from $1-8$ kpc. Our mass ranges are consistent with prior work from cluster-based samples \citep[e.g.,][]{Koda:2015aa,van-der-Burg:2016aa,Danieli:2019aa}, as well as \ion{H}{1}-selected field samples \citet{Leisman:2017aa,Kado-Fong:2021aa}. We infer that a large fraction ($\sim 70\%$) of the galaxies in this sample qualify as an ``ultra-diffuse'' galaxy by the \citet{van-Dokkum:2015aa} definition, but we also find a significant number of low-mass galaxies of $\sim 10^7$~\msun\ that fall on the ``normal'' mass-size relation. We find that the number of galaxies at a given size falls as a power-law with $n \propto r^{-3.3 \pm 0.6}$, with a slope that is similar to that reported by \citet{Prole:2021aa}. 

Our inferred number densities are consistent with known galaxy populations at $M_* \approx 10^7$~\msun, but then at higher masses become a rarer sub-constituent as they grow further and further from the mass-size ridge-line. They constitute only $10^{-3}$ of the galaxy population at $M_* \approx 10^9$~\msun\ (Figure \ref{fig:mass_size_hist}). Our mass function is surprisingly consistent with that of \ion{H}{1}-selected UDG samples for $M_* > 10^8$~\msun, suggesting that we are probing similar populations, and that star-forming and quiescent LSB galaxies have similar space densities.

A driving question as we continue to discover larger samples of LSB galaxies is whether they simply represent the smooth tail of the galaxy size distribution, or whether a distinct formation path is needed to describe them \citep[see also][]{VanNest:2021aa}. A related question is whether LSB galaxies are over-represented in rich environments as suggested by \citet[][]{Koda:2015aa} and \citet{van-der-Burg:2016aa}, although there is some evidence that the number density of LSB galaxies may rise in lower-mass groups \citep{Mancera-Pina:2018aa}. 

At the highest masses in our sample ($M_* > 10^{8}$~\msun), we do find tentative evidence that the LSB galaxies constitute an excess at large size compared to the default assumption of a log-normal distribution \citep[see also][]{Danieli:2019aa}. At $M_* \approx 10^{9}$~\msun, the galaxy population is just beginning to bifurcate into blue disks and red compact systems \citep[e.g.,][]{Lange:2015aa}. The observed excess of red LSB galaxies at this mass scale may be somehow related to this transition. Alternatively, there may be such an excess of large outliers from the mass-size ridge-line at all masses that our survey is simply not sensitive enough to detect. In this case, the question still remains whether the extreme sizes are providing hints about the dark matter halo masses \citep[e.g.,][]{Kravtsov:2013aa} and/or spin distributions \citep[e.g.,][]{Dalcanton:1997aa,Amorisco:2016aa}. Detailed modeling of kinematics for more LSB galaxies may help make progress on this point \citep[e.g.,][]{Greco:2018ac,Wasserman:2019aa,Mancera-Pina:2020aa}. 

By cross-matching our redder galaxies with groups and clusters, we also find that the galaxies projected near groups do tend to fall towards the more extended and more massive end of our sample. However, only 16\% of the red galaxies are associated with $M_h > 10^{13}$~\msun\ groups. Furthermore, the cross-correlation of red galaxies excluding the cluster-associated LSB galaxies still shows signal at $z>0.1$, suggesting that not all of the UDG-like galaxies in our sample can be associated with rich clusters. They may, however, all be satellites. Only a small handful of ``isolated'' and ``red'' LSB galaxies are currently known \citep[e.g.,][]{Martinez-Delgado:2016aa,Polzin:2021aa}, most of which are very near to the Local Group and may have tidally interacted with it \citep[e.g.,][]{Teyssier:2012aa}. \citet{Prole:2021ab} estimate that 25\% of the isolated LSB population are red.

We are currently in the process of repeating the \citet{Greco:2018aa} search on the full HSC data set. This will represent a factor of four increase in area, and thus a substantive improvement in statistical power. At that time, we will more fully model the expected mass and redshift distribution to address the question of whether these LSBs are true mass-size outliers. We will also be in a position to divide the sample into finer color and size bins. The upcoming Vera Rubin Observatory Legacy Survey of Space and Time \citep{LSST:2009} will provide data of comparable depth over much wider areas, with interesting new prospects for cross-correlation. At the same time, we anticipate deriving direct distances via surface brightness fluctuations \citep[e.g.,][]{Greco:2021} to a number of the nearest galaxies (within 15 Mpc) of the new sample, to complement the radial velocities currently available from gas \citep{Karunakaran:2020aa}. These larger samples, with multi-wavelength data, will address the key factors that lead galaxies to be LSB, whether they be internal or environmental. These surveys will enable a full census of dwarf galaxies in this mass and surface brightness range.

\section{acknowledgements}
JPG thanks Yi-Kuan Chiang for valuable discussions about cross correlation redshifts. We thank the anonymous referee for excellent and timely comments. JEG is supported in part by NSF grant AAG-1007052. SD is supported by NASA through Hubble Fellowship grant HST-HF2-51454.001-A awarded by the Space Telescope Science Institute, which is operated by the Association of Universities for Research in Astronomy, Incorporated, under NASA contract NAS5-26555. 

\bibliographystyle{aasjournal}
\bibliography{Greco_et_al_2018.bib}

\appendix

\section{Validation with GAMA}
\label{sec:Gamatest}

%%%%%%%%%%%%%%%%%%%%%%%%%%%%%%%%%%
\begin{figure*}
\begin{minipage}{.98\textwidth}
    \centering
    \includegraphics[width=0.4\textwidth]{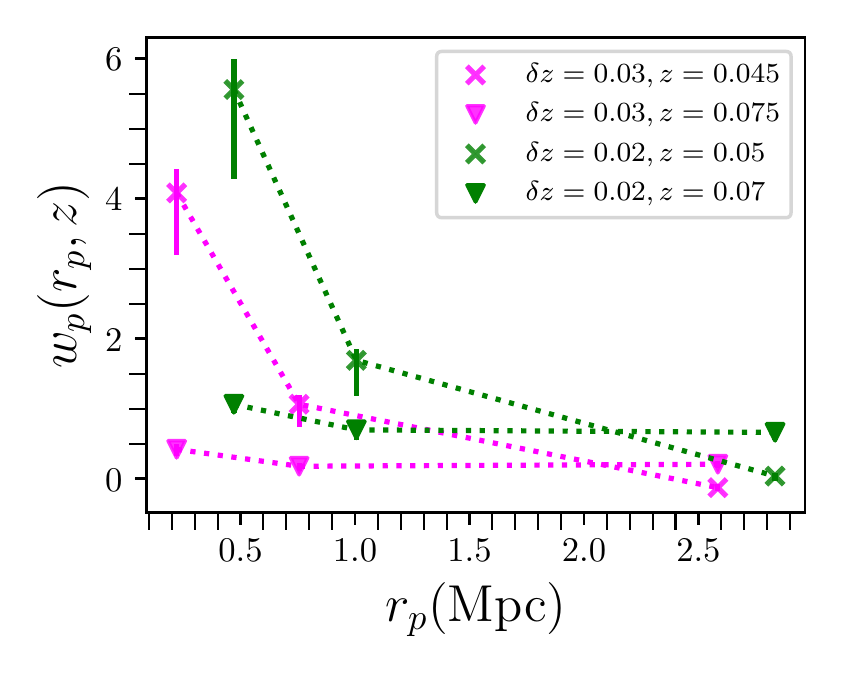}
    \includegraphics[width=0.4\textwidth]{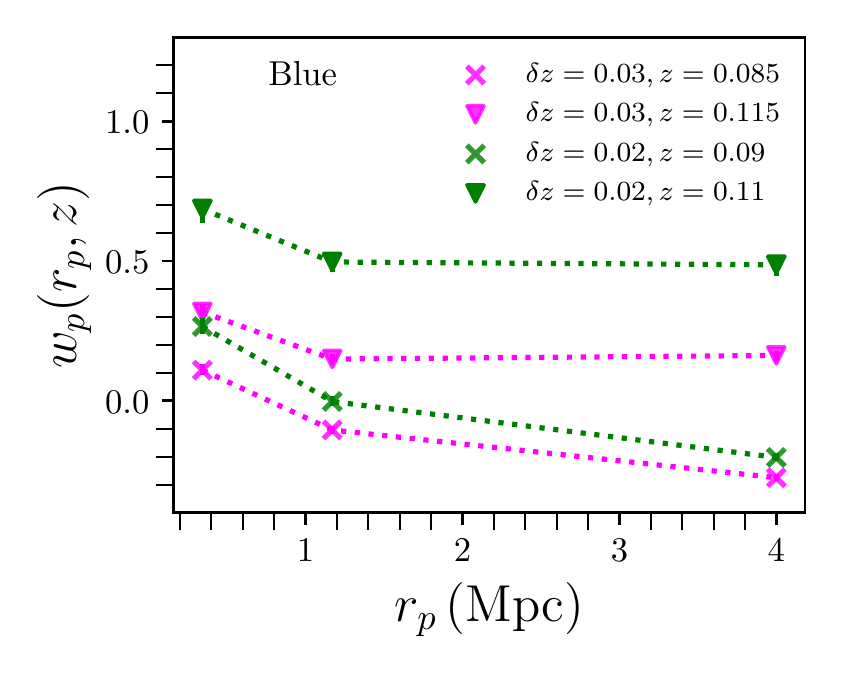}
\end{minipage}
\begin{minipage}{.98\textwidth}
    \centering
    \hskip -2mm
    \includegraphics[width=0.42\textwidth]{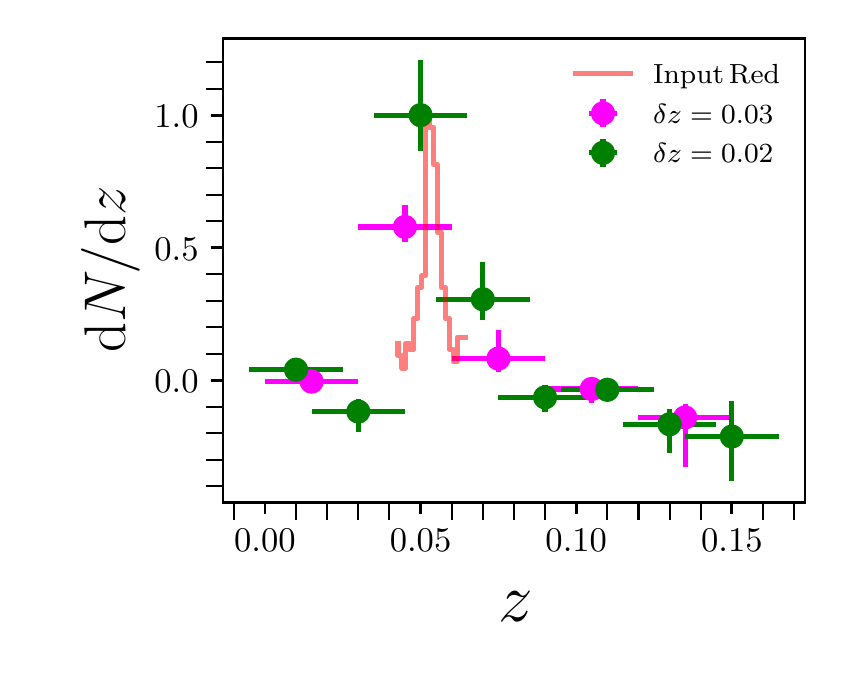}
    \includegraphics[width=0.42\textwidth]{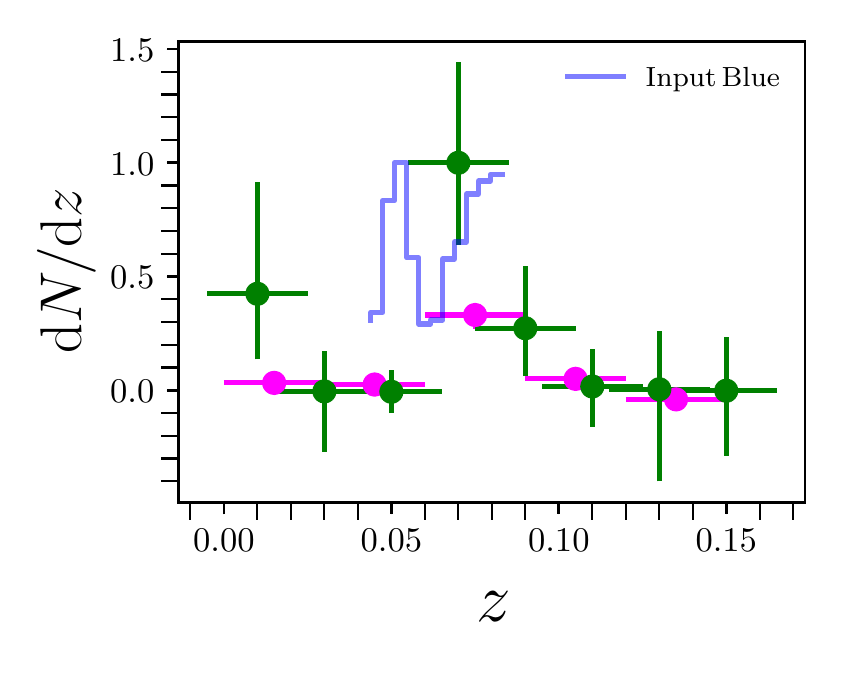}
\end{minipage}
\caption{{\it Top}: Cross-correlation for red (left) and blue (right) GAMA test galaxies, with $18<M_g<19$ mag. {\it Bottom}: Resulting inferred redshift distributions for red (left) and blue (right) populations, with the input distribution shown in histogram. Regardless of binning, we see a clear signal in the red galaxies at $z \sim 0.05$, in accord with the input redshift distribution. The situation is more challenging for the blue subset, but the mean redshift in the distribution still does match the input distribution. 
\label{fig:wbargamalow}}
\end{figure*}
%%%%%%%%%%%%%%%%%%%%%%%%%%%%%%%%%%

We present a test to gain some intuition about how the technique will work for our low redshift sample over a limited volume. 

We construct a test sample using the GAMA survey \citep{Driver:2009aa}, selecting galaxies with known redshifts to test our ability to recover the input redshift distribution \citep[see a more sophisticated analysis in][]{Schmidt:2013aa}. We use the NASA-SDSS Atlas (NSA) as the tracer sample, and to maximize spatial overlap we focus only on the GAMA equatorial fields G09, G12, and G15, for roughly 180 deg$^2$, which is comparable to our HSC area of 200 deg$^2$. Recall that we do not have masks or random catalogs for the NSA, but the GAMA team do provide random catalogs \citep{Farrow:2015aa}.  

From the GAMA survey, we select a sample of low-mass galaxies with $18<r<19$~mag and $0.04<z<0.065$ and we start with red ($g-r > 0.8$) galaxies for a total of 255 galaxies). We restrict our attention to a narrow range in magnitude to mimic our galaxies of interest. The input redshift distribution is basically a delta function (Fig.\ \ref{fig:wbargamalow}). We find that we can recover the mean redshift of $\langle z \rangle = 0.054$ with $\langle z_{\rm out} \rangle = 0.052 \pm 0.015$. We cannot recover the comparable redshift distribution for the blue galaxies, so we widen the redshift distribution to $0.04<z<0.085$ and for 730 blue ($g-r < 0.5$) galaxies we find $\langle z_{\rm out} \rangle = 0.064 \pm 0.014$ when the truth is $\langle z \rangle = 0.066$.

These tests demonstrate that when we input narrow redshift signals, even at very low redshift, we are able to recover them. We do not generate spurious signals at higher redshift. These tests give us some confidence that our result, namely the apparently high-redshift tail to the LSB galaxy distribution, is unlikely to be an artifact of our methodology.

\section{Green Galaxies}

\begin{figure*}
\begin{minipage}{.98\textwidth}
    \centering
    \includegraphics[width=0.3\textwidth]{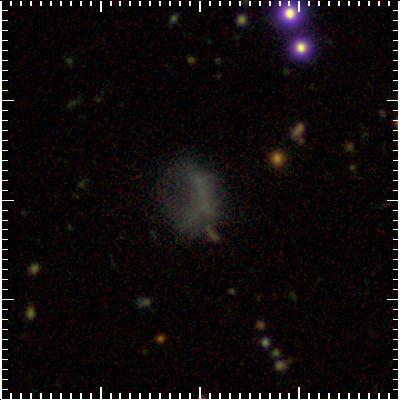}
    \includegraphics[width=0.3\textwidth]{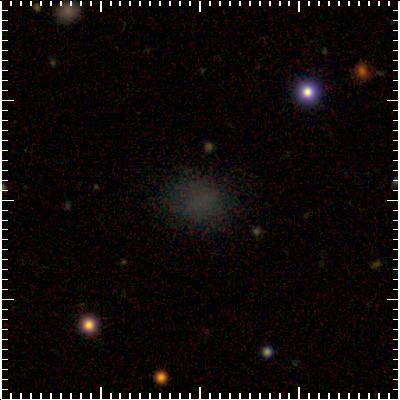}
    \includegraphics[width=0.3\textwidth]{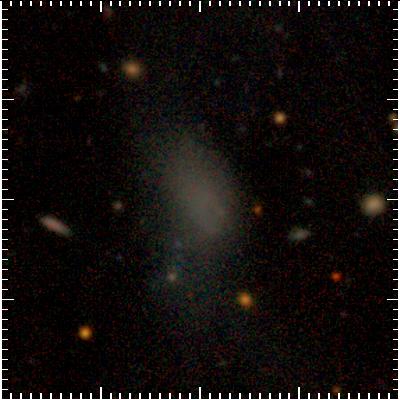}
\end{minipage}
\begin{minipage}{.98\textwidth}
    \centering
    %\hskip -2mm
    \includegraphics[width=0.3\textwidth]{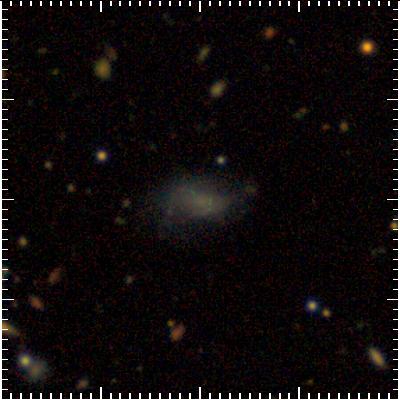}
    \includegraphics[width=0.3\textwidth]{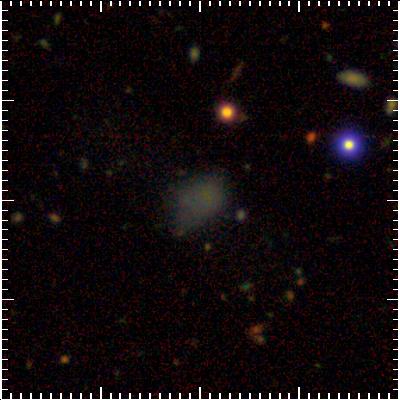}
    \includegraphics[width=0.3\textwidth]{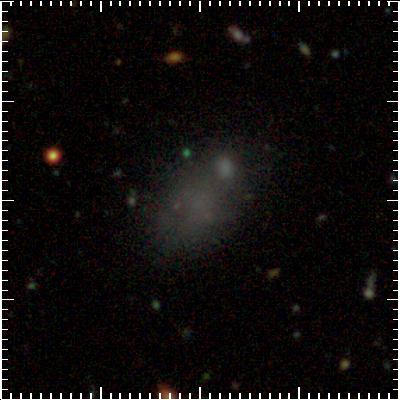}
\end{minipage}
\vskip 3mm
\caption{Examples of ``green'' galaxies that are considered to be in the main sample, to be compared with the rejected examples in the next figure. Images are 24\arcsec\ on a side.
\label{fig:greenkept}}
\end{figure*}

\begin{figure*}
\begin{minipage}{.98\textwidth}
    \centering
    \includegraphics[width=0.3\textwidth]{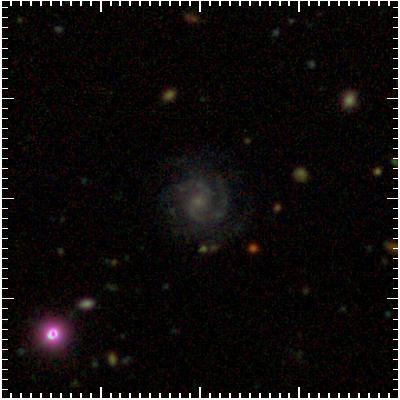}
    \includegraphics[width=0.3\textwidth]{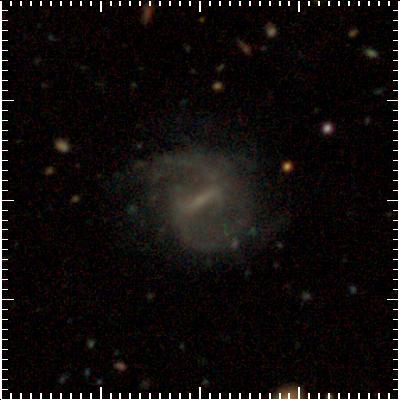}
    \includegraphics[width=0.3\textwidth]{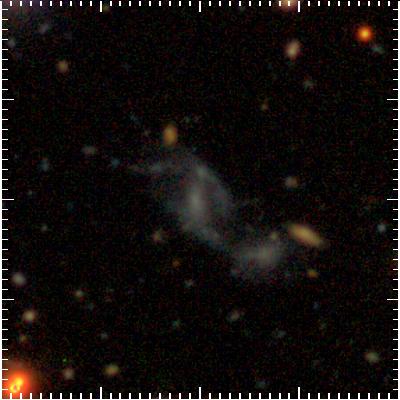}
\end{minipage}
\begin{minipage}{.98\textwidth}
    \centering
    %\hskip -2mm
    \includegraphics[width=0.3\textwidth]{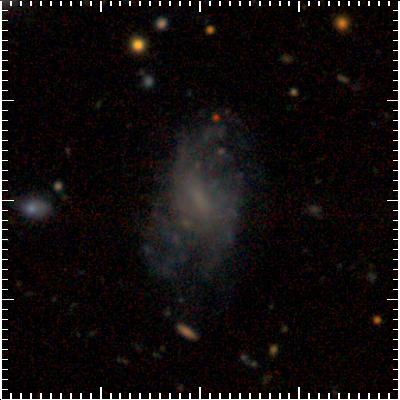}
    \includegraphics[width=0.3\textwidth]{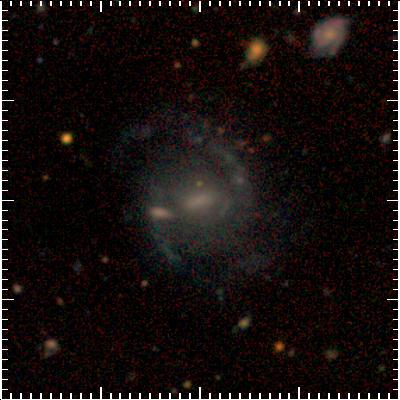}
    \includegraphics[width=0.3\textwidth]{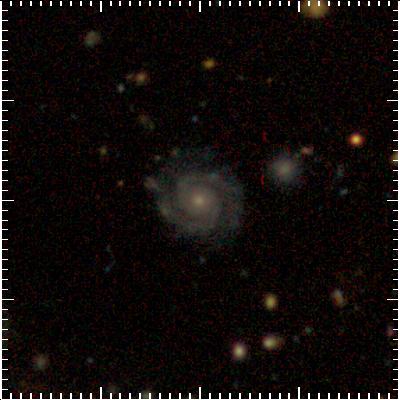}
\end{minipage}
\vskip 3mm
\caption{Examples of ``green'' galaxies that we flag as likely background. We remove them on the basis of an apparent redder bulge-like component along with clear spiral structure. Images are 24\arcsec\ on a side.
\label{fig:rejected}}
\end{figure*}

Here we present examples of green galaxies deemed to be in the sample (Figure \ref{fig:greenkept}) and green galaxies deemed to be in the background (Figure \ref{fig:rejected}).

%%%%%%%%%%%%%%%%%%%%%%%%%%%%%%%%%%
%%%%%%%%%%%%%%%%%%%%%%%%%%%%%%%%%%

%%%%%%%%%%%%%%%%%%%%%%%%%%%%%%%%%%
%%%%%%%%%%%%%%%%%%%%%%%%%%%%%%%%%%

\section{SDSS DR12 as the reference population}

We look at the cross-correlation signal with the LSB galaxies and galaxies in the SDSS DR12 large-scale structure catalog \citep{Reid:2016aa}. In this case, we do know the spatial mask, so we can construct random catalogs for the SDSS DR12 sample. We see no evidence of correlation beyond $z \sim 0.15$. 

%%%%%%%%%%%%%%%%%%%%%%%%%%%%%%%%%%
\begin{figure*}
\centering
\begin{minipage}{.98\textwidth}
    \centering
    \includegraphics[width=0.3\linewidth]{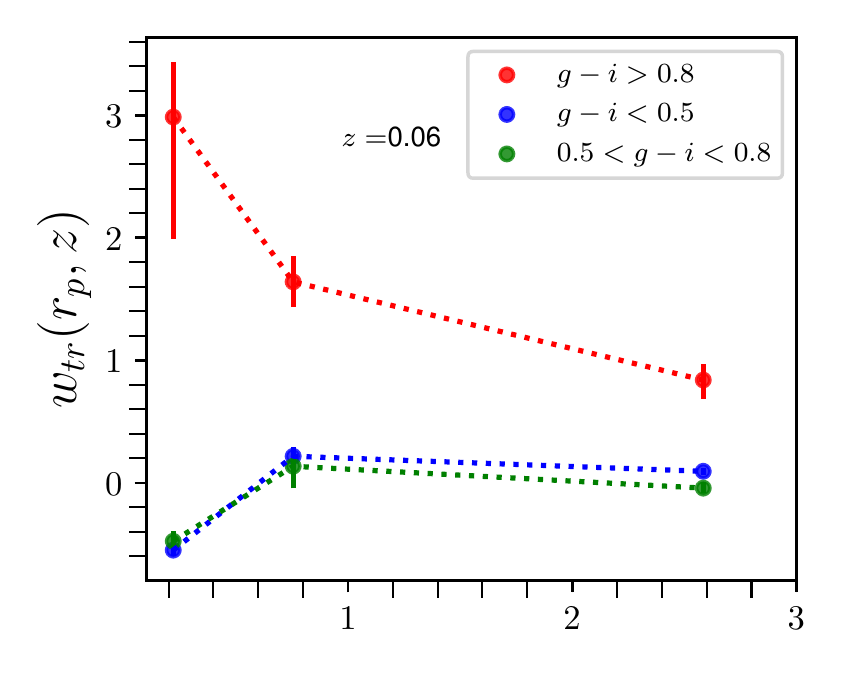}
    \includegraphics[width=0.3\linewidth]{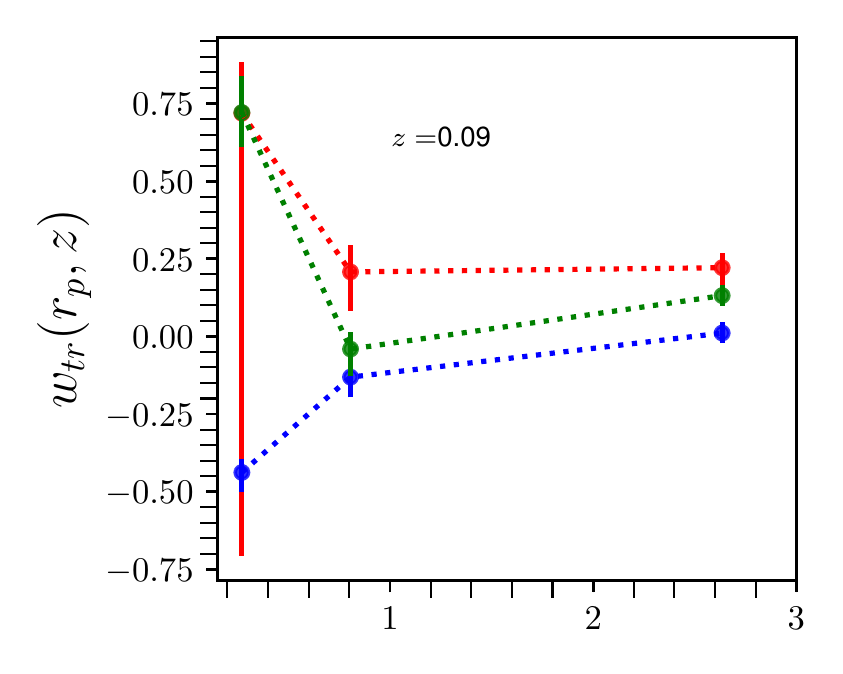}
    \includegraphics[width=0.3\linewidth]{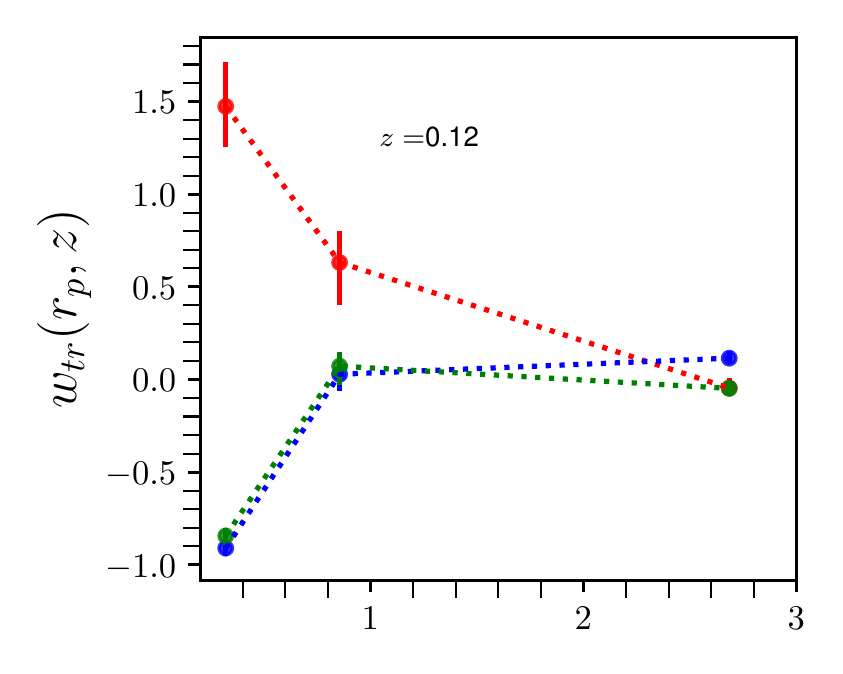}
\end{minipage}
\begin{minipage}{.98\textwidth}
    \centering
    \includegraphics[width=0.3\linewidth]{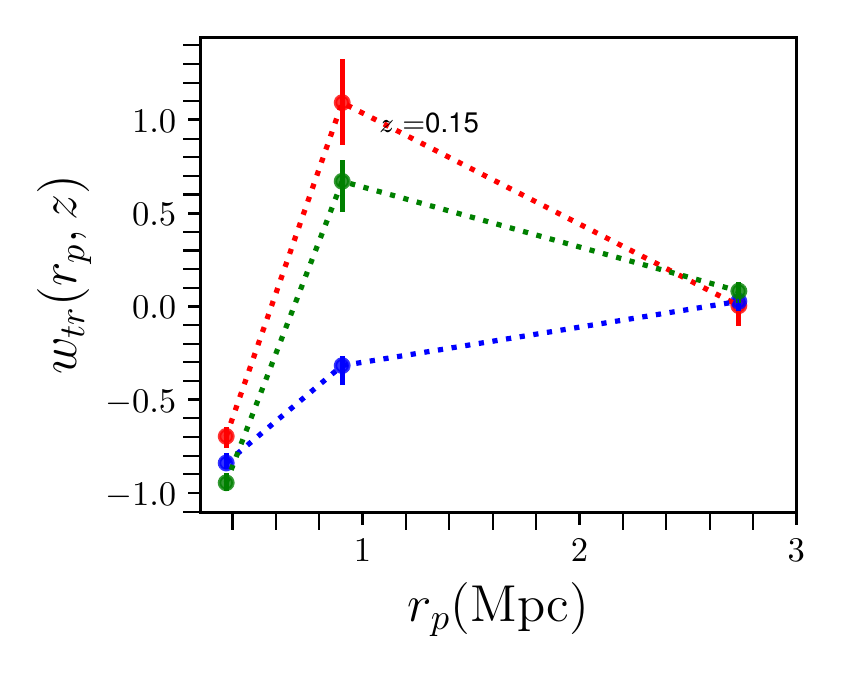}
    \includegraphics[width=0.3\linewidth]{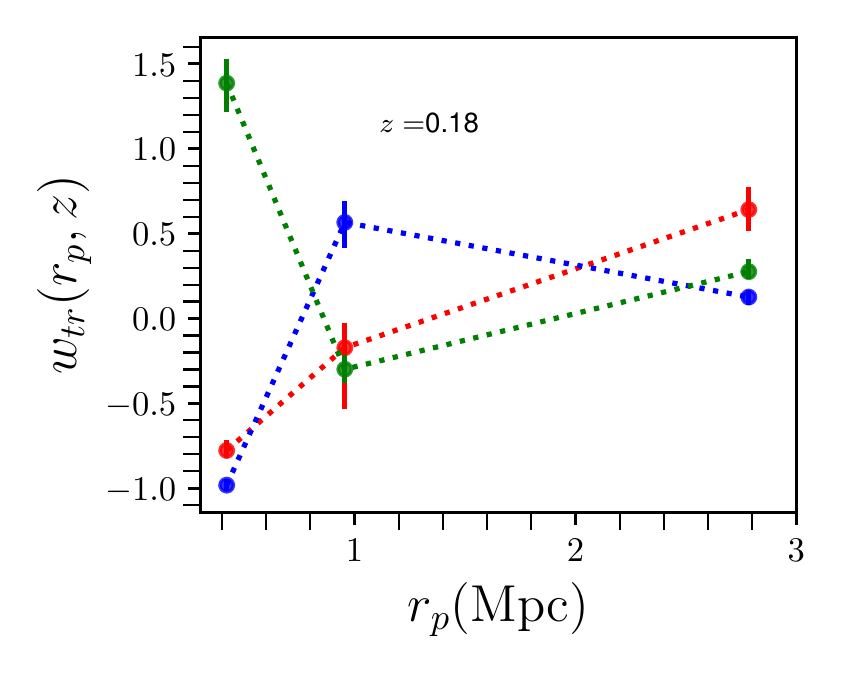}
\end{minipage}
\caption{The cross-correlation signal between the LSB galaxies and the SDSS DR12 large-scale structure catalog. We see a clear correlation with the red galaxies but not the green or blue at $z > 0.04$. 
\label{fig:xcorcolorSDSS}}
\end{figure*}
%%%%%%%%%%%%%%%%%%%%%%%%%%%%%%%%%%

\end{document}